\documentclass[prd,aps,showpacs,floats,floatfix,showpacs,superscriptaddress,twocolumn,nofootinbib]{revtex4}

\usepackage{psfrag}
\usepackage[dvips]{graphicx}
\usepackage{amsmath}
\usepackage{amsfonts}
\usepackage{color}
\usepackage{hyperref}

\newcommand{\myremark}[1]{}

\newcommand{\beq}{\begin{equation}}
\newcommand{\eeq}{\end{equation}}
\newcommand{\bea}{\begin{eqnarray}}
\newcommand{\eea}{\end{eqnarray}}
\newcommand{\ba}{\begin{array}}
\newcommand{\ea}{\end{array}}
\newcommand{\pp}[2]{\frac{\partial #1}{\partial #2}}
\newcommand{\bigip}[2]{\Big\langle #1, #2 \Big\rangle}

\newcommand{\hL}{\hat{\mathbf{L}}}
\newcommand{\hS}{\hat{\mathbf{S}}}

\newcommand{\ve}{\mathbf{e}}
\newcommand{\vT}{\mathbf{T}}
\newcommand{\vJ}{\mathbf{J}}
\newcommand{\vS}{\mathbf{S}}
\newcommand{\vL}{\mathbf{L}}

\begin{document}

\title{Detecting gravitational waves from precessing binaries of spinning compact
objects. II. Search implementation for low-mass binaries}

\author{Alessandra Buonanno}

\affiliation{Laboratoire AstroParticule et Cosmologie (APC), 11 place Marcelin Berthelot, 75005 Paris, France.}

\altaffiliation{UMR 7164 (CNRS, Universit\'e Paris 7, CEA, Observatoire de Paris).
Current address: Institut d'Astrophysique de Paris, 98$^{\rm bis}$ boulevard Arago,
75013 Paris, France.}

\author{Yanbei Chen}

\affiliation{Max-Planck-Institut f\"ur Gravitationsphysik, Albert-Einstein-Institut, Am M\"uhlenberg 1, D-14476 Golm bei Potsdam,
Germany}

\author{Yi Pan}

\affiliation{Theoretical Astrophysics and Relativity, California
Institute of Technology, Pasadena, CA 91125}

\author{Hideyuki Tagoshi}

\affiliation{Department of Earth and Space Science, Graduate School of
Science, Osaka University, Toyonaka 560-0043, Japan}

\author{Michele Vallisneri}

\affiliation{Jet Propulsion Laboratory,
California Institute of Technology, Pasadena, CA 91109}

\begin{abstract}
Detection template families (DTFs) are built to capture the essential features of true gravitational waveforms using a small set of phenomenological waveform parameters. Buonanno, Chen, and Vallisneri [Phys.\ Rev.\ D \textbf{67}, 104025 (2003)] proposed the ``BCV2'' DTF to perform computationally efficient searches for signals from precessing binaries of compact stellar objects.
Here we test the signal-matching performance of the BCV2 DTF for asymmetric--mass-ratio binaries, and specifically for double--black-hole binaries with component masses $(m_1,m_2) \in [6,12]M_\odot \times [1,3]M_\odot$, and for black-hole--neutron-star binaries with component masses $(m_1,m_2) = (10 M_\odot, 1.4 M_\odot)$; we take all black holes to be maximally spinning.
We find a satisfactory signal-matching performance, with fitting factors averaging between 0.94 and 0.98. We also scope out the region
of BCV2 parameters needed for a template-based search, we evaluate
the template match metric, we discuss a template-placement strategy, and we estimate the number of templates needed for searches at the LIGO design sensitivity.
In addition, after gaining more insight in the dynamics of spin--orbit precession, we propose a modification of the BCV2 DTF that is parametrized by \emph{physical} (rather than phenomenological) parameters. We test this modified ``BCV2P'' DTF for the $(10 M_\odot, 1.4 M_\odot)$ black-hole--neutron-star system, finding a signal-matching performance comparable to the BCV2 DTF, and a reliable parameter-estimation capability for target-binary quantities such as the chirp mass and the opening angle (the angle between the black-hole spin and the orbital angular momentum).
\end{abstract}
\date{August 15, 2005}
\maketitle

\section{Introduction}
\label{sec:intro}

As ground-based gravitational-wave (GW) detectors based on laser interferometry \cite{gwdet} approach their design sensitivities, the emphasis in data analysis is shifting from upper-limit studies \cite{ligo} to proper detection searches. In addition, the length of data-taking runs is stretching to several months, with typical duty cycles approaching unity, substantiating the need for \emph{online} (or at least \emph{real-time}) searches to be performed as data become available. It is then crucial to develop search algorithms that maximize the number of detections while making efficient use of computational resources.

Inspiraling binaries of black holes (BHs) and/or neutron stars (NSs) are among the most promising \cite{burgay} and best-understood sources for GW interferometers, which can observe the waveforms emitted during the adiabatic phase of these inspirals, well described by post-Newtonian (PN) calculations \cite{2PN}. For these signals, the search algorithms of choice are based on \emph{matched filtering} \cite{matched}, whereby the detector output is compared (\emph{i.e.}, correlated, after noise weighting) with a \emph{bank} of theoretically derived signal \emph{templates}, which encompass the GW signals expected from systems with a prescribed range of physical parameters.

Reference \cite{bcv1} introduced the phrase ``detection template
families'' (DTFs) to denote families of signals that capture the
essential features of the true waveforms, but depend on a smaller
number of parameters, either physical or phenomenological
(\emph{i.e.}, describing the waveforms rather than the sources). At
their best, DTFs can reduce computational requirements while achieving
essentially the same detection performance as exact templates; however, they are less adequate for upper-limit studies, because they may
include nonphysical signal shapes that result in increased
noise-induced triggers, and for parameter estimation, because the
mapping between template and binary parameters may not be
one-to-one, or may magnify errors. In Ref.\ \cite{bcv1}, the ``BCV1''
DTF was designed to span the families of nominally exact (but
partially inconsistent inspiraling-binary waveforms obtained
using different resummation schemes to integrate the PN equations.

A reduction in the number of waveform parameters is especially necessary when the binary components carry significant spins not aligned with the orbital angular momentum; spin--orbit and spin--spin couplings can then induce a strong precession of the orbital plane, and therefore a substantial modulation of GW amplitude and phase \cite{ACST94}. Detection-efficient search templates must account for these effects of spin, but a straightforward parametrization of search templates by the physical parameters that can affect precession results in intractably large template banks.

To solve this problem, several DTFs for precessing compact binaries have been proposed in the past decade \cite{ACST94,apostolatos0,apostolatos1,apostolatos2,GKV,GK,Gpc}. A DTF based on the \emph{Apostolatos ansatz} for the evolution of precession frequency was amply investigated in Refs.\ \cite{GKV,Gpc}, and an improved version was proposed in Ref.\ \cite{GK}. However, the computational resources required by the Apostolatos-type families are still prohibitive; more important, their signal matching performance (\emph{i.e.}, the fitting-factor FF) is not very satisfactory.

In Ref.\ \cite{bcv2}, Buonanno, Chen, and Vallisneri analyzed the physics of spinning-binary precession and waveform generation, and showed that the modulational effects can be isolated in the evolution of the GW polarization tensors, which are combined with the detector's \emph{antenna patterns} to yield its response. As a result, the response of the detector can be written as the product of a carrier signal and a complex modulation factor; the latter can be viewed an extension of the Apostolatos formula. In Ref.\ \cite{bcv2}, the precessing waveforms were cast into a mathematical form (the linear combination of three simpler signals, with complex coefficients) that allows searching automatically and economically over all the precession-related parameters, except for a single parameter $\mathcal{B}$ that describes the timescale of modulation. Henceforth, we shall refer to the template family proposed in Ref.\ \cite{bcv2} as the ``BCV2'' DTF.

In Ref.\ \cite{bcv2}, the BCV2 DTF was tested for precessing BH--BH
binaries with high total mass ($12 M_\odot < M < 30 M_\odot$) and
comparable component masses, and for the single mass configuration
$(10 + 1.4) M_\odot$, representative of NS--BH systems. In all cases,
the signal-matching performance was good (FF $> 0.9$), with consistent
improvements over search templates that do not include precessional effects (for
instance, in the NS--BH system the FF increases from $\sim 0.78$ to
$\sim 0.93$). Signals from precessing binaries with asymmetric
component masses are harder to match, because they have more orbital
and precessional cycles (\emph{i.e.}, more complex waveforms) in the
band of good interferometer sensitivity.

In this paper, we extend the BCV2 performance analysis of Ref.\
\cite{bcv2} to asymmetric mass ratios, taking into consideration
systems with component masses $(m_1,m_2) \in [6,12]M_\odot \times
[1,3]M_\odot$, for which we expect a large number of precession cycles (see Fig.\ \ref{fig:cycles} below). In addition, we estimate the region of the DTF parameter space that must be included in a search for such systems; we
calculate the template match metric \cite{BSD,O,OS99}; we provide a
strategy for template placement; last, we estimate the number of
templates required for the search.  After reconsidering the
Apostolatos ansatz, we are also able to shed new light on the
phenomenological parameter $\mathcal{B}$ that describes the timescale
of modulation; indeed, we derive an explicit formula for the evolution
of the precession angle in terms of the physical parameters of the
binary, and we use this formula to propose a modification of the BCV2
DTF that dispenses with $\mathcal{B}$.

While this paper is concerned with DTFs for precessing binaries, we note that a \emph{physical} template family for \emph{single-spin} precessing compact binaries was proposed in Ref.\ \cite{bcv2}, and thoroughly tested in Ref.~\cite{pbcv1}. The attribute ``physical'' is warranted because the family is obtained by integrating the PN equations \cite{2PN} in the time domain, and the templates are labeled by the physical parameters of the binary. Furthermore, Ref.\ \cite{pbcv2} showed that the single-spin physical family has a satisfactory signal-matching performance also for the waveforms emitted by \emph{double-spin} precessing compact binaries, at least for component masses $(m_1,m_2) \in [3,15] M_\odot \times [3,15] M_\odot$; moreover, the parameters of the best-fit single-spin templates can be used to estimate the parameters of the double-spin target systems \cite{pbcv2}. However, this physical template family may be more complicated to implement and more computationally expensive (and therefore less attractive for use in online searches) than the frequency-domain DTFs such as BCV2.

This paper is organized as follows. In Sec.~\ref{sec:apos}, we briefly review the BCV2 DTF and the Apostolatos ansatz, and we discuss how the phenomenological parameter $\mathcal{B}$, which describes the timescale of precession, can be related to the physical parameters of the binary.
In Sec.~\ref{sec:nonphysDTF}, we discuss the signal-matching performance of the BCV2 DTF over a range of binary component masses. In Sec.~\ref{sec:physDTF}, we introduce a version of the BCV2 DTF modified to include the physical evolution of the precession angle in single-spin binaries, and we test its performance for NS--BH inspirals. In Sec.~\ref{sec:metric}, we compute the template match metric for the BCV2 DTF. In Sec.~\ref{count}, we provide a strategy for template placement, and we estimate the number of templates required in a search. Last, in Sec.\ \ref{conc} we summarize our conclusions.

In the following, the binary component masses are denoted by $m_1$ and $m_2$ (with $m_1 > m_2$); the symmetric mass ratio and the total mass by $\eta = m_1 m_2/M^2$ and $M=m_1 + m_2$; the binary component spins by $\vS_1 =\chi_1 \,m_1^2$ and $\vS_2 =\chi_2 \,m_2^2$. For single-spin binaries, we assume $\vS_1 =\chi \,m_1^2$ and $\vS_2 =0$. Throughout the paper, the signal-matching performance of DTFs is evaluated against a target model for precessing binaries governed by Eqs.~(6)--(32) of Ref.~\cite{pbcv1}; this target model is valid in the adiabatic phase of inspiral, when dynamics are correctly described by PN equations. We use an analytic fit to the LIGO-I design noise spectrum (given, \emph{e.g.}, by Eq.\ (28) of Ref.~\cite{bcv1}); we adopt the standard formalism of matched-filtering GW detection; we follow the conventions of Ref.~\cite{pbcv2}, which contains a useful glossary of matched-filtering notions and quantities; last, we always set $G = c = 1$.

\section{Features of precession dynamics in single-spin binaries}
\label{sec:apos}

\subsection{Review of the Apostolatos ansatz and of the BCV2 DTF}
\label{sec:aposreview}

Apostolatos, Cutler, Sussman, and Thorne (ACST) \cite{ACST94} investigated
orbital precession in binaries of spinning compact objects in two special cases:
(i) equal-mass binaries ($m_1=m_2$), where the spin--spin coupling is switched off, and (ii) single-spin binaries ($S_2=0$). In these cases, precessional dynamics can always be categorized as \emph{simple precession} or \emph{transitional precession}. In simple precession, the direction of the total angular momentum $\hat{\vJ}$ is roughly constant, while the orbital angular momentum $\vL$ and the total spin $\vS=\vS_1+\vS_2$ precess around it. ACST were able to derive an analytical solution
for the evolution of simple precession (see Sec. IV of Ref.~\cite{ACST94}).
Transitional precession occurs when, during evolution, $\vL$ and $\vS$ have roughly the same amplitude and become nearly antialigned. When this happens, $\left|\vJ\right|$ is almost zero and $\hat{\vJ}$ can change suddenly and dramatically. Although transitional precession is too complicated for analytical treatment, it occurs rarely \cite{bcv2,ACST94}, so we will ignore it in this paper.

GW signals from generic precessing binaries are well approximated by simple-precession waveforms when the ACST assumptions are valid, which happens for two classes of binaries: (i) BH--BH binaries with comparable component masses where the spin--spin
interaction can be neglected, which are equivalent to systems where a single object carries the total spin of the system; (ii) BH--NS or BH--BH binaries with very asymmetric mass ratios, which can be approximated as single-spin systems because the spin of the lighter object is necessarily small. It is not guaranteed \emph{a priori} that simple-precession waveforms can describe also signals emitted by BH--BH binaries with intermediate mass ratios and/or important spin--spin effects. However, it was recently shown \cite{pbcv2} that simple-precession waveforms are adequate also for these classes of binaries, although the dynamical evolution of $\vL$ and $\vS$ can exhibit rather different features \cite{apostolatos1}.
\begin{figure}
\begin{center}
\includegraphics[width=0.4\textwidth]{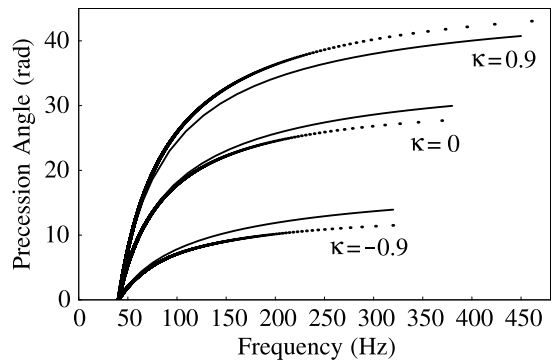}
\caption{Evolution of precession angle (radians) as a function of GW frequency, 
for $(10+1.4)M_{\odot}$ binaries with opening angles $\kappa=0.9, 0$ and $-0.9$.
The dotted curves show the numerical evolution of our target systems, while the continuous curves follow the analytical expression~\eqref{Omega-p}.
\myremark{[MV: this figure could be lightened by reducing the number of points in the dotted curves.]}
\label{fig:alpha-p}}
\end{center}
\end{figure}

In simple precession, $\hL$ and $\hS$ precess around $\hat{\vJ}$ (which is roughly fixed) with the precession frequency given by Eq.\ (42) of Ref.\ \cite{ACST94},
\beq
\label{omegap}
\Omega_p\equiv\frac{d\alpha_p}{dt}
=\left(2+\frac{3m_2}{2m_1}\right)\frac{J}{r^3}\,.
\eeq
ACST found that the evolution of the precession angle $\alpha_p$ can be approximated by power laws in two extreme cases.  When $L \gg S$ and $J\simeq L$, using leading-order Newtonian expressions, we find
\begin{equation}
\label{NewtLr}
\begin{gathered}
L=\eta M^2 (M \omega)^{-1/3}\,, \qquad
r=\displaystyle\left(\frac{M}{\omega^2}\right)^{1/3}\,, \\
\frac{\dot{\omega}}{\omega^2} = \frac{96}{5} \eta \,(M\omega)^{5/3} \,,
\end{gathered}
\end{equation}
where $\omega$ is the orbital angular frequency, and $r$ is the orbital separation, so it is straightforward to get $\alpha_p=\mathcal{B}_1 f^{-1}$; this regime corresponds to comparable-mass binaries, or to binaries at large separations (\emph{i.e.}, in the early stages of the inspiral). When $L\ll S$ and $J\simeq S$, we have $\alpha_p=\mathcal{B}_2 f^{-2/3}$; this regime corresponds to binaries with large mass asymmetry, or to binaries at small separations (\emph{i.e.}, in the late stages of the inspiral). The analytical expressions for the coefficients $\mathcal{B}_1$
and $\mathcal{B}_2$ are given by Eq.~(45) of Ref.\ \cite{ACST94}, and depend only on the masses and on the total spin of the binary, but \emph{not} on the opening angle between the spin and the orbital angular momentum. Although the power laws were derived for simple precession under the ACST assumptions, it turns out that they can model the dynamics of more general configurations, as shown in Refs.~\cite{bcv2,pbcv2}.

On the basis of the ACST analysis, Apostolatos \cite{apostolatos1} reasoned that GWs from precessing binaries should be modulated by the orbital precession frequency
$\Omega_p$, and suggested adding an $\Omega_p$-dependent modulation term to the nonspinning waveform phasing,
\begin{equation}
\label{eq:apansatz}
\psi_{\rm spinning}\rightarrow\psi_{\rm nonspinning}+{\mathcal
  C}\cos(\delta+{\mathcal B}f^{-2/3})
\end{equation}
(this is the Apostolatos ansatz). Although the resulting DTF has higher fitting factors with precessing-binary waveforms than nonspinning DTFs, it is not completely satisfactory \cite{apostolatos1, GKV}, especially because of the huge computational cost implicit in adding the three parameters $\mathcal{C}, \delta, \mathcal{B}$,
which are \emph{intrinsic} (\emph{i.e.}, they increase the dimensionality of search template banks).

In Ref.~\cite{bcv2}, Buonanno, Chen, and Vallisneri proposed a DTF (BCV2) based on a modification of Eq.\ \eqref{eq:apansatz},
\begin{widetext}
\begin{multline}
\label{bcv2}
h(\psi_0,\psi_{3/2},{\mathcal B},f_{\rm cut},{\mathcal C}_k;f)= \\
f^{-7/6} \left[({\mathcal C}_1+i{\mathcal C}_2)+({\mathcal C}_3+i{\mathcal
    C}_4)\cos({\mathcal B}f^{-2/3})
+({\mathcal C}_5+i{\mathcal C}_6)\sin({\mathcal B}f^{-2/3})\right]
\theta(f_{\rm cut}-f)e^{2\pi ift_0} \times
e^{i[\psi_0f^{-5/3}+\psi_{3/2}f^{-2/3}]}\,;
\end{multline}
\end{widetext}
here $t_0$ is the signal's time of arrival, and $f_\mathrm{cut}$ is the cutoff frequency;
precessional effects are modeled by the modulation terms
$\cos({\mathcal B}f^{-2/3})$ and $\sin({\mathcal B}f^{-2/3})$ in
the complex amplitude, separately from the nonspinning evolution of phase.
Possible modulation morphologies are enriched by the presence of the complex linear-combination coefficients ${\mathcal C}_3+i{\mathcal C}_4$ and ${\mathcal C}_5+i{\mathcal C}_6$, improving the efficiency of matching to target waveforms. Indeed the complex modulation terms modulate both amplitude and phase of the nonspinning signal.
An important feature of the BCV2 DTF is that little computational cost is added by the search over the parameters $\mathcal{C}_{1,\ldots,6}$, which are \emph{extrinsic} (\emph{i.e.}, the detection statistic can be maximized analytically over them).

Although the $f^{-2/3}$ power law adopted in the BCV2 DTF (\ref{bcv2})
is expected to be valid only for $S\gg L$, (\emph{i.e.}, for binaries with asymmetric
mass ratios, or in the late stages of inspiral), the DTF yields high fitting factors also for comparable-mass binaries, as verified in Ref.\ \cite{bcv2} by Monte Carlo simulation. The reason is probably to be found in the broad variety of modulation morphologies parametrized by the $\mathcal{B}$ and $\mathcal{C}_{1,\ldots,6}$ parameters.

\subsection{Analysis of the DTF parameter $\mathcal{B}$}
\label{sec:beta}

In this section we relate the phenomenological precession parameter $\mathcal{B}$ to the physical parameters of the binary (the two masses, $m_1, m_2$, the cosine between the directions of the total spin and the orbital angular momentum, $\kappa \equiv \hat{\vL}\cdot\hat{\vS}$, and the magnitude of the spin, $\chi \equiv S_1/m_1^2$). In doing so, we clarify why the BCV2 DTF is capable of mimicking the precessional effects in the target signal.

In Fig.\ 16 of Ref.\ \cite{bcv2}, the distribution of the best-fit DTF $\mathcal{B}$ is plotted against the target-system parameter $\kappa$. The target waveforms were generated at 2PN order, for BH--NS binaries of component masses $(10+1.4)M_{\odot}$ with maximally spinning BHs. The spread of the data points corresponds to uniform distributions of the initial spin and angular-momentum orientations.
The points seem to cluster around three lines, but no explanation is offered for this interesting feature. We are now able to explain this behavior; what we learn in the process will enable us to construct an improved DTF (BCV2P) parametrized by physical parameters (see Sec.\ \ref{sec:physDTF}).


Although the best-fit DTF parameter $\mathcal{B}$ is not, strictly speaking,
physical, it is clearly related to the evolution of the precession angle $\alpha_p$ in the target system. Moreover, we expect the best-fit $\mathcal{B}$ to be a function of the target-system opening angle $\kappa$ (as seen in Fig.\ 16 of Ref.\ \cite{bcv2}), except in the limits $L \gg S$ and $L \ll S$, where the power laws $\alpha_p=\mathcal{B}_{1}f^{-1,-2/3}$ do not include $\kappa$.
Let us see what function we should expect. From Eq.~(49) of \cite{ACST94}, we have
\beq
\label{Omega-p}
\Omega_p=\left(2+\frac{3m_2}{2m_1}\right)
\sqrt{1+2\kappa\gamma+\gamma^2}\frac{L}{r^3}\,,
\eeq
which was obtained by expressing the total angular momentum $J$ in Eq.~(\ref{omegap})
in terms of the orbital angular momentum $L$. In Eq.~(\ref{Omega-p}),
$\gamma(t)$ denotes the quantity $S/L(t)$; the dependence of $\Omega_p$ on $\kappa$ vanishes with $\gamma \gg 1$ (\emph{i.e.}, with $L \ll S$) or $\gamma \ll 1$ (\emph{i.e.}, with $L \gg S$). Using the leading-order Newtonian expressions for
$L$ and $r$ given in Eq.~\eqref{NewtLr}, we can integrate $\Omega_p$ analytically and obtain $\alpha_p$ as a function of $m_1$, $m_2$, $\kappa$, and $\chi$:
\begin{multline}
\alpha_p^{N}(f) = \frac{5}{384} \frac{4m_1+3m_2}{m_1} \times \\
\Bigl\{
-A \, \left[ (2 - 3\kappa^2) \, \chi_M^2 + \kappa \, \chi_M \, v^{-1} + 2 \, v^{-2} \right] \\
+ 3 \, \kappa \, (1 - \kappa^2) \, \chi_M^3 \, \log \left[ \kappa \, \chi_M + v^{-1} + A \right] 
\Bigr\} + \mathrm{const.}
\label{alphapk}
\end{multline}
where
\begin{equation}
\begin{gathered}
v = (\pi M f)^{1/3}\,, \qquad
\chi_M=\frac{m_1}{m_2} \,\chi\,, \\
A=\sqrt{\chi_M^2 + 2 \, \kappa \, \chi_M \, v^{-1} + v^{-2}}\,,
\end{gathered}
\end{equation}
and where $f=\omega/\pi$. The ``$N$'' in $\alpha_p^N$ stands for ``Newtonian''. These expressions are equivalent to Eqs.\ (63a) and (63b) of Ref.~\cite{ACST94}.
Note that an analytical expression including higher PN corrections could also be given. However, for simplicity we prefer to restrict ourselves to the lowest order.
In Fig.~\ref{fig:alpha-p}, for a binary of mass $(10+1.4)M_{\odot}$
and for several values of $\kappa$, we compare the analytical precession angles $\alpha^N_p(f)$ with numerical values obtained from our target models by projecting  $\hat{\vL}(f)$ onto the plane perpendicular to the vector $\hat{\vJ}$ (which is constant because we only consider simple precession), and recording the cumulative angle swept by the projected image. We see that our leading-order formula reproduces the shape of the numerically-obtained curve, although the quantitative difference is appreciable. This is due to the fact that we write $\alpha_p^N$ at the Newtonian order; the agreement would otherwise be perfect.
\begin{figure}
\begin{center}
\includegraphics[width=0.45\textwidth]{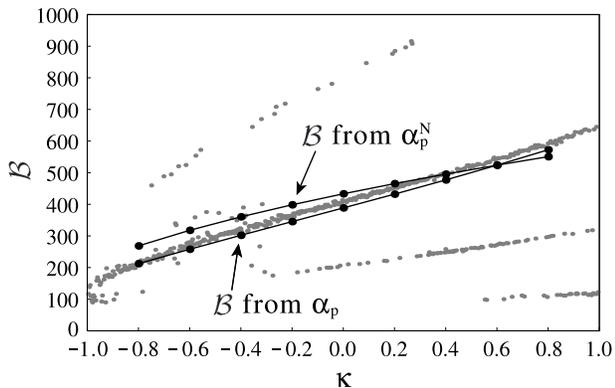}
\caption{
Best-fit values of the BCV2 DTF parameter $\mathcal{B}$ as a function of the target-system opening angle $\kappa$, for $(10+1.4)M_{\odot}$ binaries with uniformly distributed initial orientations of spin and angular momentum.
The larger connected dots show the equivalent $\mathcal{B}$ as evaluated from
Eq.\ \eqref{beta150} using the numerical $\alpha_p(f)$ and the analytical
$\alpha_p^N(f)$, respectively. \label{fig:betakappa}}
\end{center}
\end{figure}

We can now try to explain the dependence of the DTF parameter $\mathcal{B}$ on the 
parameters of the binary, and the clustering seen in Fig.\ 16 of Ref.~\cite{bcv2}.
Since the power laws $\mathcal{B}f^{-2/3,1}$ cannot match $\alpha_p(f)$ exactly,
we establish a correspondence by requiring that the instantaneous rate of change of the two precession angles be equal at the approximate frequency of best detector sensitivity, $\sim 150\,{\rm Hz}$ (appropriate for initial LIGO):
\beq
\label{eq:equivalentf}
\frac{d}{dt} \mathcal{B}f^{-2/3} \equiv \Omega_p^t = \Omega_p^s \equiv \frac{d}{dt} \alpha_p(t)\,, \quad f=150 \, {\rm Hz} \,,
\eeq
and therefore
\beq
\label{beta150}
\mathcal{B}=
-\frac{3}{2}f^{5/3}\left.\frac{d\alpha_p(f)}{df}\right|_{f=150Hz}\,.
\eeq
Here $\Omega_p^t$ denotes the template's equivalent precession frequency, while $\Omega_p^s$ denotes the target's orbital precession frequency at 150 Hz.
Figure \ref{fig:betakappa} shows a test of the correspondence, 
performed for $(10+1.4)M_{\odot}$ binaries with uniformly distributed values of $\kappa$ and of the initial orientations of total spin and orbital angular momentum.
The small dots indicate the best-fit values of $\mathcal{B}$, similarly
to Fig.\ 16 of Ref.\ \cite{bcv2}. The larger dots mark the pairs $(\kappa,\mathcal{B}(\kappa))$ obtained from Eq.~\eqref{beta150} using the analytical $\alpha_p^N(f)$ and the numerical $\alpha_p(f)$, respectively. 
The pairs line up quite well with the linear cluster in the middle of Fig.\ \ref{fig:betakappa}, which includes the majority of data points. Thus, it is correct to state that the value of $\mathcal{B}$ represents the rate of change $\Omega_p^t =\Omega_p^s$ of the precession angle in the middle of the frequency band of good detectory sensitivity. Similar conclusions hold also for binaries
with masses $(m_1,m_2) \in [6,12]M_\odot \times [1,3]M_\odot$.
The other linear clusters seen in Fig.~\ref{fig:betakappa}  
correspond roughly to $2\Omega_p^t= \Omega_p^s$ and $\Omega_p^t=2\Omega_p^s$, for reasons explained in the next section (the clusters in the lower corners, on the other hand, correspond to systems where the effects of precession on the waveforms are negligible).

[Note that there are some differences between Fig.~\ref{fig:betakappa} and
Fig.\ 16 of Ref.\ \cite{bcv2}.\footnote{In Fig.\ 16 of \cite{bcv2}, the $x$ axis refers to $\kappa_{\rm eff} / \kappa_{\rm eff}^{\rm max}$, where $\kappa_{\rm eff} = \hat{\vL}\cdot \vS_{\rm eff}/M^2$, with $\vS_{\rm eff}$
defined by Eq.\ (7) of Ref.\ \cite{bcv2}. In the limit in which
only one of the two bodies carries spin, the quantity $\kappa$
used in this paper equals $\kappa_{\rm eff} / \kappa_{\rm eff}^{\rm max}$.}
The linear cluster identified with $\Omega_p^t=\Omega_p^s$ in Fig.~\ref{fig:betakappa} 
corresponds in fact to the top linear cluster in Fig.\ 16 of Ref.\ \cite{bcv2}, and
the fraction of points in this cluster is significantly larger here than there.
Moreover, the top linear cluster corresponding to $\Omega_p^t=2\Omega_p^s$
cannot be seen clearly in Fig.\ 16 of Ref.\ \cite{bcv2}. These differences
are due to the adoption, for this paper, of a better numerical overlap-maximization scheme, which has a better chance of finding the true global maximum overlap.  We shall discuss the maximization scheme in detail in Sec.~\ref{sec:nonphysDTF}.]

\subsection{Higher harmonics in templates and signals}
\label{subsub}
\begin{figure*}
\begin{center}
\includegraphics[width=0.8\textwidth]{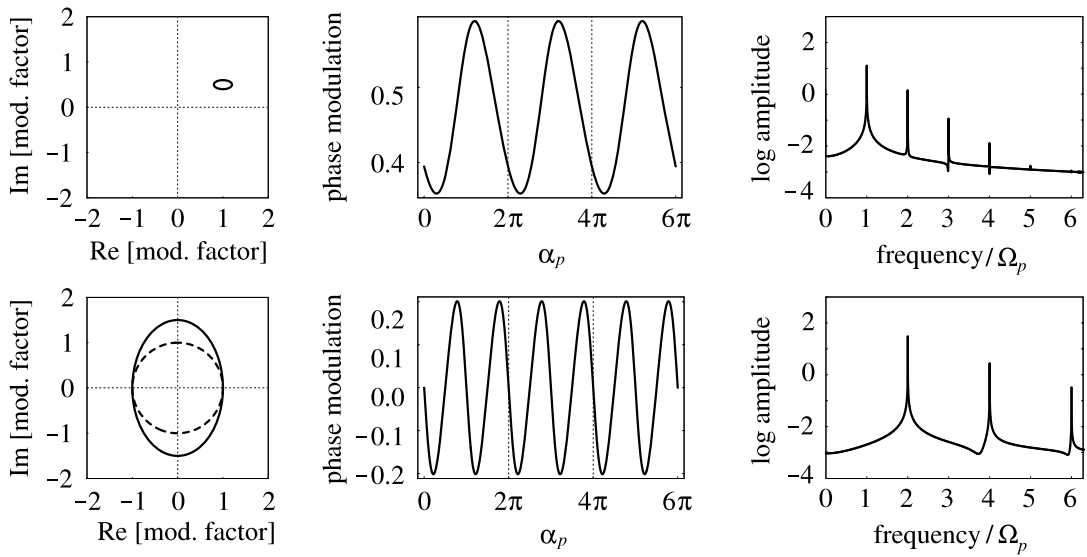}
\caption{
BCV2 DTF complex modulation factor, as given by Eq.\ \eqref{camp2}.
The top (bottom) row refers to a $\mathcal{C}_{1,2}$-dominated ($\mathcal{A}_{r,i}$-dominated) choice of the modulation parameters.
The left column shows the complex-plane trajectory of the
modulation factor when the precession angle $\alpha_p$ varies between $0$ and $2\pi$; the middle column shows the oscillatory part of phase as function of
$\alpha_p$; the right column shows the log-amplitude of the Fourier transform of phase.
\label{fig:fmod}}
\end{center}
\end{figure*}
We shall now consider why multiple clusters appear in
Fig.\ \ref{fig:betakappa}. We shall see that the precession frequencies 
$\Omega_p^t$ and $\Omega_p^s$ defined in Eq.\ \eqref{eq:equivalentf} are not usually the only modulation frequencies to appear in the spectra of the template and signal waveforms---indeed, they may not even be the dominant frequencies. It is then conceivable that if different harmonics of the precession frequency dominate in the template and signal waveforms, the maximum overlaps may occur when $\Omega_p^t$ and $\Omega_p^s$ are not equal, but instead related by integer factors.

We first consider the frequency content of the BCV2 template phase modulations. 
As discussed above,
their precession angle is $\alpha_p \equiv \mathcal{B}f^{-2/3}$, and the precession frequency $\Omega_p^t$ is given by the time derivative of $\alpha_p$. Modulation effects are included by way of the complex factor
\beq
\label{camp1}
({\mathcal C}_1+i{\mathcal C}_2)+({\mathcal C}_3+i{\mathcal
    C}_4)\cos({\mathcal B}f^{-2/3})
+({\mathcal C}_5+i{\mathcal C}_6)\sin({\mathcal B}f^{-2/3})
\eeq
(with $\mathcal{C}_{1,\ldots,6} \in \mathbb{R}$), which is
clearly a periodic function of $\alpha_p$, but is obviously 
far from a simple sinusoid with a single frequency. Recasting the factor as
\beq
\label{camp2}
({\mathcal C}_1+i{\mathcal C}_2)+
{\mathcal A}_r\cos({\mathcal B}f^{-2/3}+\varphi_r)+
i{\mathcal A}_i\cos({\mathcal B}f^{-2/3}+\varphi_i)
\eeq
(with $\mathcal{C}_{1,2}, \mathcal{A}_{r,i}, \varphi_{r,i} \in \mathbb{R}$), we see that it traces an ellipse in the complex plane as 
$\alpha_p$ varies from $0$ to $2\pi$. The shape of the ellipse is determined by
$\mathcal{A}_{r,i}$ and $\varphi_{r,i}$, and the displacement of its center from the origin by $\mathcal{C}_1+i\mathcal{C}_2$.

For two choices of the modulation parameters $\mathcal{C}_{1,2}$, $\mathcal{A}_{r,i}$, and $\varphi_{r,i}$, Fig.\ \ref{fig:fmod} shows the complex-plane trajectory of the modulation factor (left panels), the oscillatory part\footnote{That is, the residual obtained after fitting the total phase to the nonspinning phase $\phi_0+2\pi t_0f+\psi_0f^{-5/3}+\psi_{3/2}f^{-2/3}$.} of the phase as a function of $\alpha_p \equiv \mathcal{B}f^{-2/3}$ (middle panels), and the amplitude of its Fourier transform.
In the first example, we choose ${\mathcal C}_{1,2}$ to dominate, setting
${\mathcal C}_1 = 1$, ${\mathcal C}_2 = 0.5$, $\mathcal{A}_r = 0.2$,
$\mathcal{A}_i = 0.1$, $\varphi_r = 0$, and $\varphi_i = \pi/2$.
The phase is periodic in $\alpha_p$ with period $2\pi$, and it is rather close to a single sinusoid (upper middle panel). The Fourier spectrum (upper right panel) shows that the dominant template modulation frequency $\Omega_m^t$ is the precession frequency $\Omega_p^t$, and that the contributions from higher harmonics are at least an order of magnitude smaller.

In the second example, we choose ${\mathcal A}_{r,i}$ to dominate, setting
${\mathcal C}_1 = 0$, ${\mathcal C}_2 = 0$, $\mathcal{A}_r = 1$,
$\mathcal{A}_i = 1.5$, $\varphi_r = 0$, and $\varphi_i = \pi/2$.
In this case the modulation ellipse encloses the origin, so the phase contains a monotonic component that grows by $2\pi$ each time the modulation factor completes an orbit around the origin. It is easy to see that this monotonic component is simply $\alpha_p$. The oscillatory part of the phase can be obtained 
from Fig.\ \ref{fig:fmod} (lower left panel)
by taking the phase difference between points with the same real parts on the ellipse and on the unit circle. In this case, the phase
is (almost) periodic in $\alpha_p$ with period $\pi$,
as shown in the lower middle panel. The Fourier spectrum (lower right panel) shows clearly that the dominant frequency is $\Omega_m^t=2 \Omega_p^t$, and that there is no component at $\Omega_p^t$ or at any other odd harmonic.
\begin{figure*}
\begin{center}
\includegraphics[width=1\textwidth]{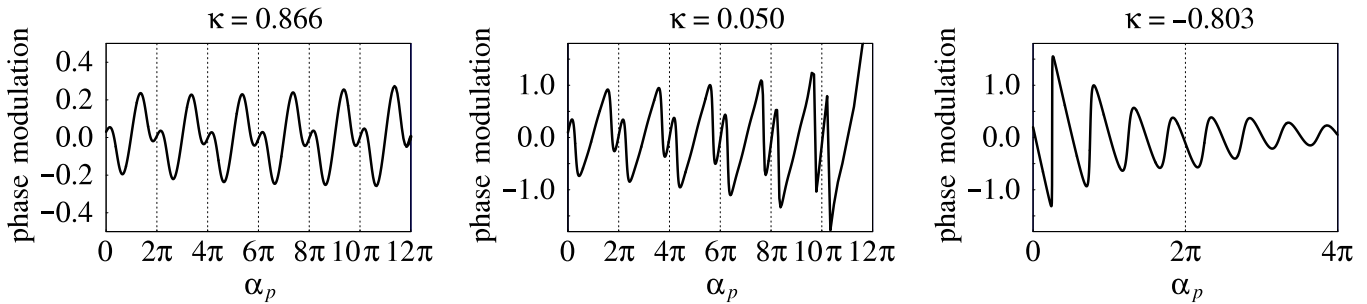}
\caption{Oscillatory part of phase modulation in target waveforms for
BH--NS binaries with $(m_1, m_2, \chi)=(10 M_\odot, 1.4 M_\odot, 1)$,
and with directional parameters such that the detector line of sight
is perpendicular to the initial $\vL$-$\vS$ plane and to the detector
plane, and that the detector is oriented along the ``$+$'' GW
polarization. The three panels correspond to $\kappa=0.866$ (for which
$\vL$ is close to $\mathbf{\Omega}_L$), $\kappa=0.050$, and
$\kappa=-0.803$ (for which $\vL$ is perpendicular to
$\mathbf{\Omega}_L$ at $f=30$ Hz).
The sharp turns observed in the second and third panels are not singularities of the phase, but happen when the projections of $\ve_+$ and $\ve_{\times}$ on the detector frame (and therefore the real and imaginary components in Eq.\ \eqref{hresp}) are both small, so that the phase can change very rapidly.
\label{fig:fmodsig}}
\end{center}
\end{figure*}

Investigating other choices of the modulation parameters, we find that
the Fourier spectra of the phase modulations have always their highest
peaks at either $\Omega_p^t$ or $2\Omega_p^t$, depending on whether
the modulation ellipse encloses the origin or not.
Most of the differences between the various cases lie in the structure of harmonics above the second. For instance, if the trajectory of the complex amplitude has very large ellipticity, higher-order harmonics can become comparable to the lowest harmonic (\emph{i.e.}, the fundamental or the second harmonic).

We now consider the frequency content of the target-waveform phase modulations. Under the \emph{precession convention} introduced in Ref.~\cite{bcv2}, the frequency-domain expression for the waveforms, at the leading order in the stationary-phase approximation, are given by
Eq.~(83) of Ref.\ \cite{bcv2}, or namely
\begin{multline}\label{hresp}
\tilde{h}_{\rm resp}(f)=-\tilde{h}_C(f)\left\{\left[\ve_+(t_f)\right]
^{jk}+i\left[\ve_{\times}(t_f)\right]^{jk}\right\} \\
\times\left(\left[\vT_+(t_f)\right]_{jk}F_+
+\left[\vT_{\times}(t_f)\right]_{jk}F_{\times}\right)
\quad\mbox{for}\;f>0\,,
\end{multline}
where $\tilde{h}_C(f)$ is the unmodulated carrier signal,
the $\vT_{+,\times}(t_f)$ are the detector polarization tensors in the
radiation frame [see Eq.(25) of BCV2], and all
precessional effects are isolated in the evolving GW polarization tensors
$\ve_{+,\times}(t_f)$. These are defined as
\bea
\ve_+&=&\ve_1\otimes\ve_1-\ve_2\otimes\ve_2\,, \nonumber \\
\ve_{\times}&=&\ve_1\otimes\ve_2+\ve_2\otimes\ve_1\,,
\eea
where $\ve_1$ and $\ve_2$ form a basis in the instantaneous orbital
plane. The time dependence of these polarization tensors enters the
waveform through terms of the form
\beq
\left[\ve_{+,\times}(t_f)\right]^{ij}
\left[\vT_{+,\times}(t_f)\right]_{ij}\,,
\label{ee}
\eeq
which can be approximated by $\mathcal{C}_{+,\times}
\cos(\mathcal{B}f^{2/3}+\delta_{+,\times})$, adopting the Apostolatos
ansatz. However, this is only an approximation, in two distinct ways.
First, the components of the polarization tensors depend quadratically on the components of the basis vectors $\ve_{1,2}$; as a consequence, even if the basis vectors contained only oscillations with frequency $\Omega_p^s$, the waveform would still end up with both $\Omega_p$ and $ 2\Omega_p$ modulations.
Second, under the precession convention, the basis vectors $\ve_{1,2}$ [and thus the quantities in Eq.\  \eqref{ee}] do not really precess together with the orbital plane around $\hat{\mathbf{J}}$ with angular velocity $\mathbf{\Omega}_L =\Omega_p^s \hat{\mathbf{J}}$; instead, they precess around the component of $\mathbf{\Omega}_L$ that is orthogonal to the orbital angular momentum $\vL$ [Eq.(72) of BCV2],
\beq
\mathbf{\Omega}_e(t)=\mathbf{\Omega}_L(t)-
\left[\mathbf{\Omega}_L(t)\cdot\hat{\vL}_N(t)\right]\hat{\vL}_N(t)\,,
\eeq
which in turn precesses together with $\vL_N(t)$ around $\mathbf{\Omega}_L(t)$. 
For this reason, the oscillations in $\ve_{1,2}$ are more complicated than simple sinusoids of frequency $\Omega_p^s$. ACST had already observed this fact, noticing that the phase modulations are certainly due to the precessional evolution of the orbital plane, but arise also from the so-called \emph{Thomas precession} term
(see Eq.~(29) of Ref.~\cite{ACST94} and the discussion around it).

In Fig.~\ref{fig:fmodsig}, we show the modulation of the target-waveform phase as a functions of $\alpha_p$ for three binary configurations. As for the discussion of template waveforms, we consider only the oscillatory part of phase modulation.
This is done here by fitting the total phase to a nonspinning phase
$\phi_0+2\pi t_0f+\psi_0f^{-5/3}+\psi_{3/2}f^{-2/3}$, and taking the residual.
In the first example on the left of Fig.\ \ref{fig:fmodsig}, the phase modulation is periodic in $\alpha_p$ with period $2 \pi$, and it is not very different from a single sinusoid. So in this case the dominant modulation frequency is the precession frequency $\Omega_p^s$. In the second example in the middle of Fig.\ \ref{fig:fmodsig}, the dominant frequency is still $\Omega_p^s$, but the $2\Omega_p^s$ component is also quite significant. In the third example on the right of Fig.\ \ref{fig:fmodsig}, the dominant frequency is $4\Omega_p^s$. Other systems with values of $\kappa$ between those used for Fig.~\ref{fig:fmodsig} show similar features,  sometimes with a larger number of frequency components.

So far, this discussion suggests that the phase-modulation frequencies $\Omega_m^t$ and $\Omega_m^s$ of the template and target waveforms are not always the
precession frequencies $\Omega_p^t$ and $\Omega_p^s$ (which depend on $\mathcal{B}$ and on the physical parameters of the target system, respectively), but can also be their integer multiples. In light of this, how can we understand the multiple clustering seen in Fig.\ \ref{fig:betakappa}? The answer is that, for a given target signal, there can be several templates whose overlap with the target is a local maximum, corresponding to different combinations of the $\Omega_p^t$ and $\Omega_p^s$ harmonics.

Suppose for example that the target waveform contains the first and second harmonics of the precession frequency, $\Omega_p^s$ and $2\Omega_p^s$; then a template with $\Omega_p^t=\Omega_p^s/2$ could match these two components with its second and fourth harmonics, a template with $\Omega_p^t=\Omega_p^s$ could match them with its first and second harmonics, while a template with $\Omega_p^t=2\Omega_p^s$ could match only the $\Omega_m^s= 2\Omega_p^s$ component with its first harmonic.  Using this reasoning, we can easily understand the existence of clusters of local maxima with $\Omega_p^t = \Omega_p^s$, $2\Omega_p^s$, $3\Omega_p^s$, \ldots, and $\Omega_p^t = \Omega_p^s/2$, $\Omega_p^s/3$, $\Omega_p^s/4$, \ldots\ However, if the DTF is able to reproduce the entire harmonic structure of the signal, then the local maximum with $\Omega_p^t=\Omega_p^s$ must also be the global maximum. The fact that in Fig.~\ref{fig:betakappa} we have clusters at $\Omega_p^t=\Omega_p^s/2$ and $\Omega_p^t=2\Omega_p^s$ suggests that the BCV2 DTF cannot do this perfectly. The analysis of the best-fit template modulation parameters for the points in the $\Omega_p^t=\Omega_p^s/2$ cluster of Fig.~\ref{fig:betakappa} (extending from $\kappa \simeq -0.2$ to $\kappa \simeq 1.0$) confirms the template frequency-doubling scenario discussed in this section.

\section{Signal-matching performance of the BCV2 and BCV2P DTFs}
\label{sec:performance}

\subsection{Performance of the BCV2 detection template family}
\label{sec:nonphysDTF}
\begin{figure*}
\begin{center}
\includegraphics[width=0.9\textwidth]{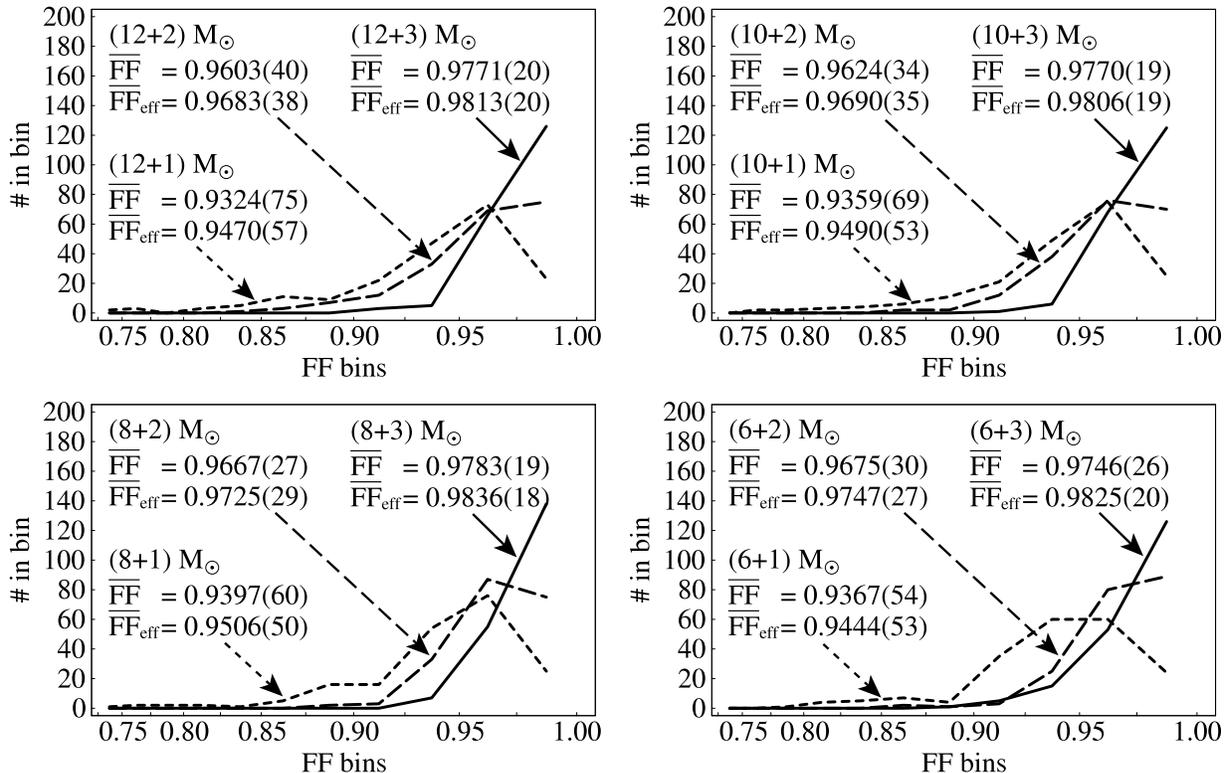}
\caption{Distribution of BCV2 DTF fitting factors for populations of target systems with component masses $(m_1,m_2)=\{6,8,10,12\}M_\odot \times \{1,2,3\}M_\odot$, and with uniformly sampled directional and local angular parameters. For each pair of masses, we include 200 target systems. The curves show the number of samples falling within each bin marked on the abscissa; the bins have equal width, but are plotted logarithmically to emphasize FFs close to unity. The figures show also the average fitting factor $\overline{\rm FF}$ and the effective average fitting factor $\overline{\rm FF}_{\rm eff}$ (as defined by Eq.\ (119) of Ref.~\cite{bcv2}), with their estimated Monte Carlo error in parentheses.
\label{fig:FFs}}
\end{center}
\end{figure*}

The basic diagnostic of DTF signal-matching performance is the \emph{fitting factor} FF ($0 \leq \mathrm{FF} \leq 1$), defined as the match between a given template in the target family and the templates in the DTF, \emph{maximized} over all the parameters of the DTF. The maximization of the match induces a many-to-one \emph{projection} between the target-signal parameters and the best-fitting DTF parameters. The way in which the maximization is carried out informs the distinction between intrinsic and extrinsic DTF parameters: full templates must be recomputed for each value of the intrinsic parameters to be explored, while the maximum over the extrinsic parameters can be computed analytically, given a choice of the intrinsic parameters. For the BCV2 DTF, the intrinsic parameters are $\psi_0$, $\psi_{3/2}$, $\mathcal{B}$, and $f_\mathrm{cut}$, and the extrinsic parameters are $\mathcal{C}_{1,\ldots,6}$ and the time of arrival $t_0$.

In our tests, the maximization of the match is performed by way of a \emph{simplex}-based search \cite{NR} in the continuous space of intrinsic parameters (as opposed to the lattice-based searches used in bank-efficiency Monte Carlos): \emph{simplex} methods have shown good efficiency in finding extrema in spaces of moderate dimensionality.
In the light of the discussion of Fig.\ \ref{fig:betakappa} in Sec.~\ref{sec:beta}, we understand that several values of the parameter $\mathcal{B}$ may yield local maxima of the match, corresponding to multiples $\Omega^s_p/2$, $\Omega^s_p$, and $2\Omega^s_p$ of the target-signal precession frequency. To improve the robustness of our search, we run it repeatedly, starting with different initial values of $\mathcal{B}$ (100, 200, 300, 400, 500, and 600, covering the likely range of $\mathcal{B}$), so the final FF is usually picked out from the best of a few local maxima. In addition, after each run we \emph{restart} the search from the current best-fit $\mathcal{B}$, creating one more chance to escape a local maximum. (By contrast, in Ref.\ \cite{bcv2} we always started simplex-based maximization at $\mathcal{B}=20$, too small compared to the values that we now see to correspond to the physical precession frequency. Because of this choice, Fig.\ 16 of Ref.\ \cite{bcv2} shows more points clustered around what are likely to be values of $\mathcal{B}$ corresponding to $\Omega^s_p/2$.)

We test the performance of the BCV2 DTF for BH--BH target systems with component masses 
\beq
\label{masses}
(m_1,m_2)=\left\{\begin{array}{c} 12M_{\odot} \\ 10M_{\odot} \\
8M_{\odot} \\ 6M_{\odot} \end{array} \right\}\times
\left\{\begin{array}{c} 3M_{\odot} \\ 2M_{\odot} \\ 1M_{\odot}
\end{array} \right\},
\eeq
as well as for BH--NS systems with $(m_1,m_2)=(10M_{\odot},1.4M_{\odot})$. We always take BHs to have maximal spins and NSs to be nonspinning. Without loss of generality, we fix the directional parameters describing GW propagation and detector orientation, while we generate randomly 200 configurations of the other directional parameters and of the local parameters of the binary (see Table I and the discussion around it in Ref.~\cite{bcv2} for the definition of directional and local parameters).
\begin{figure*}[t]
\begin{center}
\includegraphics[width=1.0\textwidth]{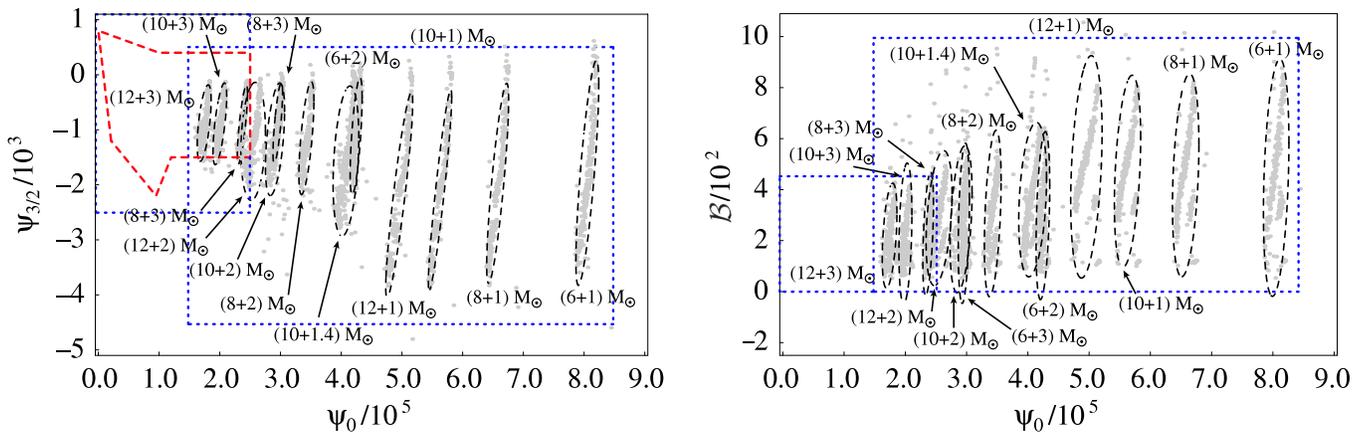}
\caption{FF projection maps onto the BCV2 intrinsic parameter space for target systems with component masses $(m_1,m_2)=\{6,8,10,12\}M_\odot \times \{1,2,3\}M_\odot$ and for $(10+1.4)M_\odot$ BH--NS systems, with uniformly sampled directional and local angular parameters. The left panel shows the $(\psi_0$,$\psi_{3/2})$ section of the map, while the right panel shows the $(\psi_0,\mathcal{B})$ section. The dots mark the values of the BCV2 parameters that achieve the maximum match for each target system. For each pair of masses, we draw an ellipse centered at the baricenter of the
corresponding dot cloud, sized to include 90\% of the dots; the axes of the ellipse follow the quadratic moments of the dots.
The dotted rectangles show the regions used in Sec.\ \ref{count} to estimate the number of BCV2 templates needed to search for the systems with component masses $(m_1,m_2) \in [6,12]M_\odot \times [1,3]M_\odot$ and $(m_1,m_2) \in [5,20]M_\odot \times [5,20]M_\odot$. In the left panel, the dashed lines enclose the region that was prescribed in Ref.\ \cite{bcv2} for the heavier systems.
\label{fig:projections}}
\end{center}
\end{figure*}

The FFs obtained are generally high, as seen by the plots of their distributions in Fig.\ \ref{fig:FFs}. That figure shows also the average fitting factor $\overline{\rm FF}$ and the effective average fitting factor $\overline{\rm FF}_{\rm eff}$, as defined by Eq.\ (119) of
Ref.~\cite{bcv2}. The FFs have a strong dependence on the mass ratio of the target binary. The more asymmetric the masses, the hardest it is to get high FFs, probably because the number of precessional cycles is larger. Due to the improved search procedure, we find slightly better FFs for $(10+1.4)M_{\odot}$ BH--NS binaries than the values obtained in Ref.\ \cite{bcv2} (see Fig.\ \ref{fig:FFs1014}). Moreover, we see that in Fig.~\ref{fig:betakappa} the relative number of points in the cluster corresponding to $\Omega_p^t=\Omega_p^s$ is increased with respect to Fig.\ 16 of Ref.~\cite{bcv2}.

An important question is what ranges of DTF parameters should be
included in a template bank to be used in a search for a given class
of target systems. Since all the BCV2 parameters are phenomenological,
the straightforward approach is to include templates corresponding to
the range of the FF projection maps (\emph{i.e.}, to include the
regions in the DTF parameter space where the maximum matches were
achieved, for a representative set of target
systems). Figure~\ref{fig:projections} shows the $(\psi_0,\psi_{3/2})$
and $(\psi_0,\mathcal{B})$ sections of the FF projection. The clusters
of maxima are labeled by the mass parameters of the target system; the
spread in each cluster corresponds to different choices of the
directional and local parameters of the binary.
The dotted rectangles show the regions used in Sec.\ \ref{count} to estimate the number of BCV2 templates needed to search for the systems analyzed in this paper
(with component masses $(m_1,m_2) \in [6,12]M_\odot \times [1,3]M_\odot$)
and for heavier systems (with component masses $(m_1,m_2) \in [5,20]M_\odot \times [5,20]M_\odot$). For comparison, the dashed polygon shows the $(\psi_0,\psi_{3/2})$ region suggested in Ref.\ \cite{bcv2} for the heavier systems.  
\begin{figure}
\begin{center}
\includegraphics[width=0.48\textwidth]{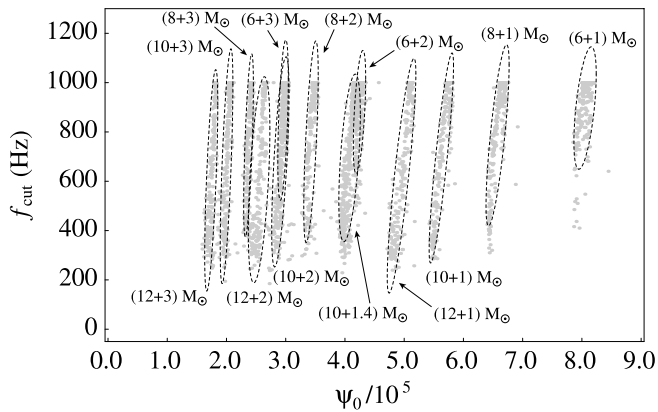}
\caption{
FF projection maps onto the BCV2 $(\psi_0,f_{\rm cut})$ parameter subspace. See Fig.\ \ref{fig:projections} for details. The cutoff seen at $1000$ Hz is due to a hard constraint arbitrarily imposed on the $f_{\rm cut}$ search range.
\label{fig:psi0fcut}}
\end{center}
\end{figure}

Figure~\ref{fig:psi0fcut} shows the $(\psi_0,f_\mathrm{cut})$ section of the projection map. The range of $f_{\rm cut}$ has a weak dependence on the masses of the target system, very probably because the target signals end their evolution at the margin of the frequency band of good interferometer sensitivity (for LIGO-I), where little signal power is lost by excluding higher frequencies. Throughout this paper, we define the target signal ending frequency as the instantaneous GW frequency at the minimum energy circular orbit (MECO), as given by Eqs.\ (11)--(12) of Ref.\ \cite{bcv2}, and plotted in Figs.\ 5 and 6 of Ref.\ \cite{bcv2}, and in Fig.\ 1 of Ref.\ \cite{pbcv1}. This frequency is inversely proportional to the total mass of the binary, and it is smaller for antialigned spin and orbital angular momentum (\emph{i.e.}, negative $\kappa$).

In fact, it seems that the $f_{\rm cut}$ parameter can be dropped altogether, as suggested by the following test: using the BCV2 DTF but \emph{fixing} $f_{\rm cut} = 400$ Hz,
we evaluate FFs for $(6+1) M_\odot$ and $(12+3) M_\odot$ systems.
The first mass configuration was chosen because, although it corresponds to the largest ending frequency (\emph{i.e.}, the smallest total mass) among the systems studied in this paper, it shows the largest distribution of $\psi_{3/2}$ in Fig.\ \ref{fig:projections}; thus, by removing $f_{\rm cut}$ from the maximization of the overlap, we could expect a very mild dependence on $f_{\rm cut}$, but a noticeable change in the FF projection.
The second mass configuration was chosen because it corresponds to the smallest ending frequency (\emph{i.e}, the largest total mass) among our systems. For both classes of systems the FF drops by \emph{only} $0.5\%$ with respect to searches that include $f_{\rm cut}$, with insignificant changes in the FF projection ranges.

\begin{figure*}
\begin{center}
\includegraphics[width=0.8\textwidth]{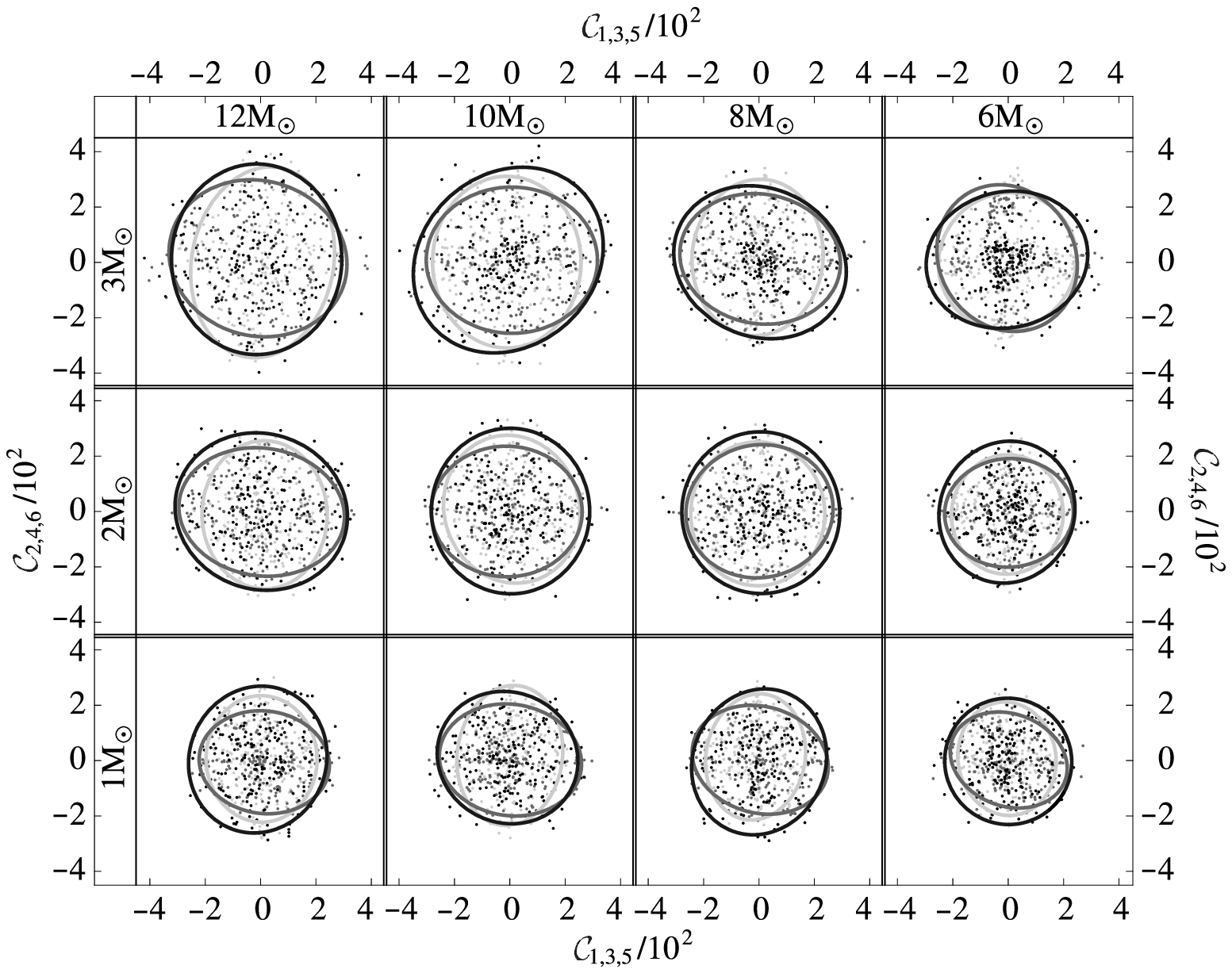}
\caption{
FF projection maps onto the BCV2 extrinsic parameter space, for the same target systems considered in Figs.\ \ref{fig:projections} and \ref{fig:psi0fcut}. Dots of different darkness mark the best-fit values of $(\mathcal{C}_1,\mathcal{C}_4)$, $(\mathcal{C}_2,\mathcal{C}_5)$, $(\mathcal{C}_3,\mathcal{C}_6)$. The ellipses are drawn as in Fig.\ \ref{fig:projections}.
\label{fig:Cprojections}}
\end{center}
\end{figure*}

Although the analytic maximization over the entire ranges of the extrinsic parameters carries little computational burden, it is useful to constrain their ranges as tightly as possible to reduce the rate of false alarms. By constraining the ranges, we are in effect reducing the range of candidate signals that are compared against the detector output, and that have a (small) chance of being triggered by detector noise alone. In Fig.~\ref{fig:Cprojections} we show the $({\mathcal C}_1,{\mathcal C}_2)$,
$({\mathcal C}_3,{\mathcal C}_4)$, and $({\mathcal C}_5,{\mathcal C}_6)$ section of the FF projection.
The absolute magnitude of the $\mathcal{C}_k$ ranges is application-dependent, since it is determined by the overall normalization of the waveforms.\footnote{We normalize all target signals and templates by assuming arbitrary luminosity distances: this has no effect on the computation of FFs and template match metrics, which always involve normalized waveforms.}
However, we notice that the $\mathcal{C}_{3,\ldots,6}$ coefficients (which govern the amplitude of modulations) have magnitude comparable to the $\mathcal{C}_{1,2}$ coefficients (which multiply the unmodulated waveform), and that the area occupied by the points shrinks slightly with decreasing total mass and more asymmetric mass ratios (corresponding to higher ending frequencies, and therefore greater signal power to be normalized).

\begin{figure}
\begin{center}
\includegraphics[width=0.42\textwidth]{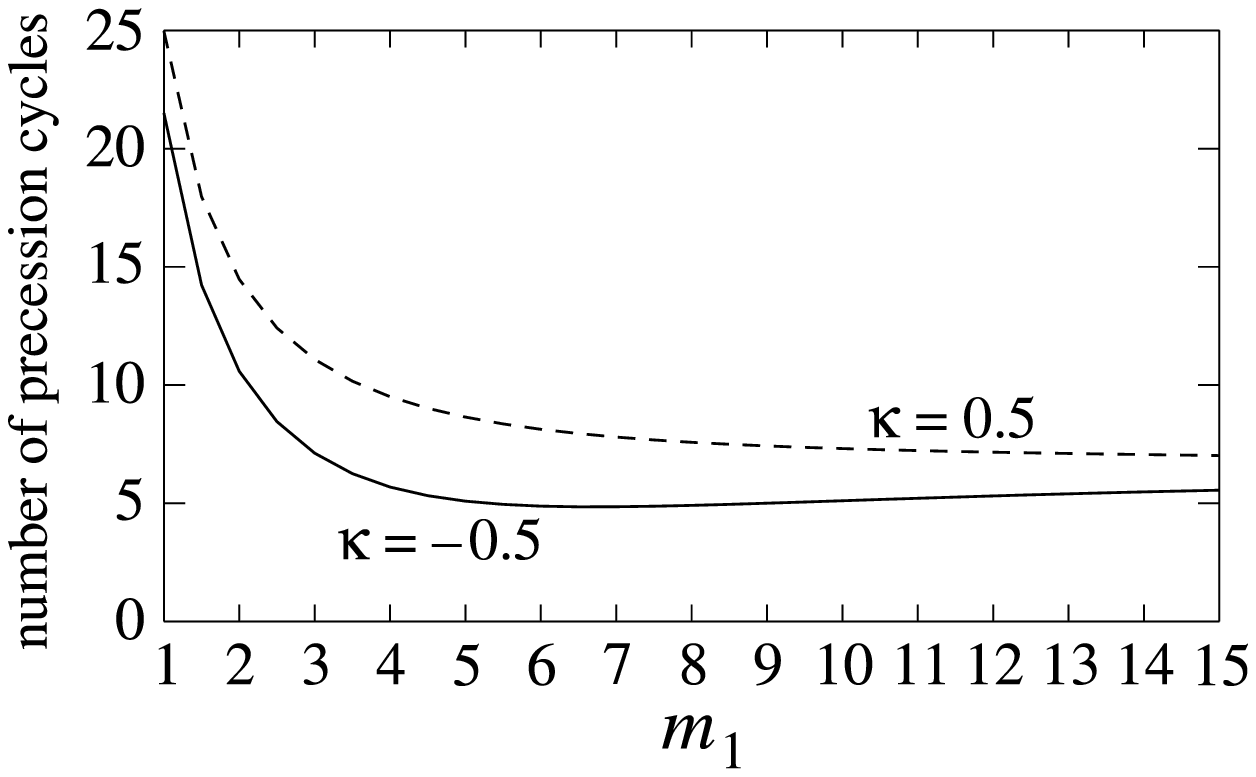}
\includegraphics[width=0.42\textwidth]{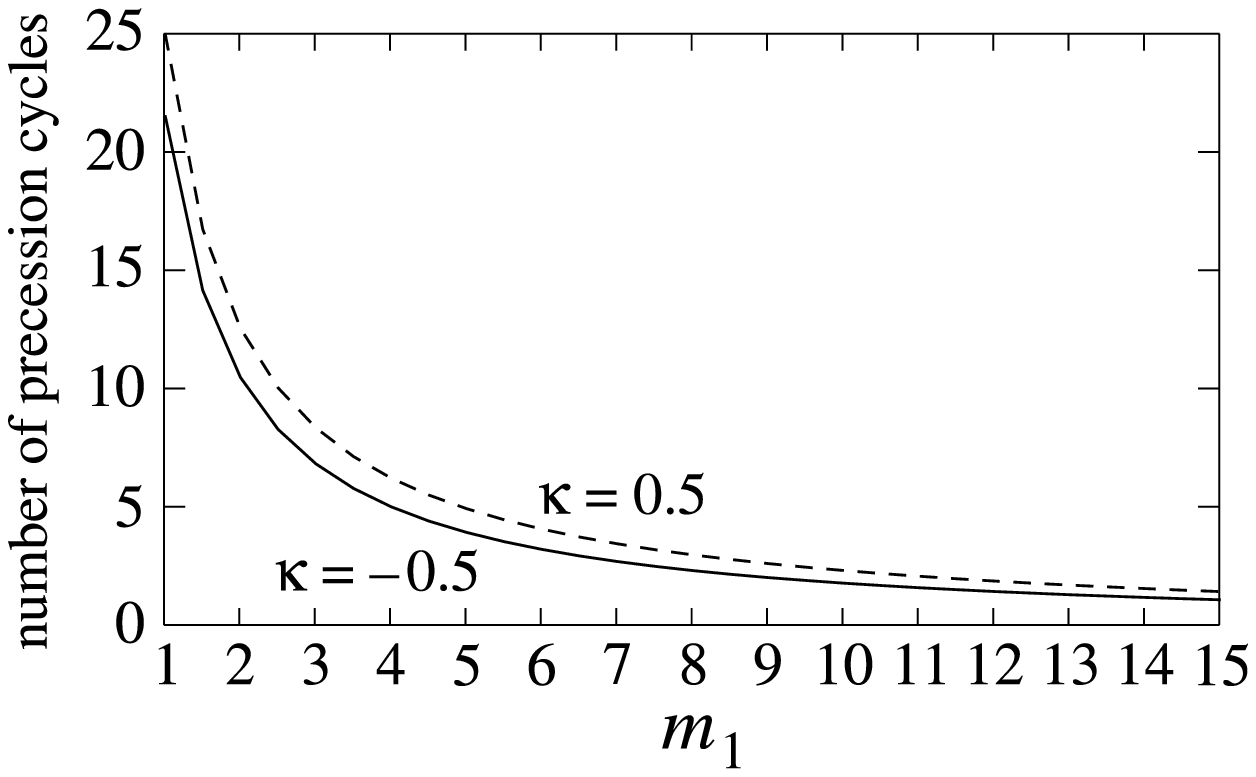}
\caption{Number of precession cycles for asymmetric--mass-ratio binaries (upper panel) with $m_2 = 1 M_\odot$, and for equal-mass binaries (lower panel), as functions of $m_1$, for two values of the opening angle $\kappa$.
\label{fig:cycles}}
\end{center}
\end{figure}

Asymmetric--mass-ratio binaries are not the only systems having a large number of precession cycles. In Fig.~\ref{fig:cycles}, we show the
number of precession cycles between 40 Hz and the GW frequency of the
test-mass ISCO, as evaluated from Eq.\ (\ref{alphapk}). In the upper panel, 
we consider asymmetric--mass-ratio binaries with $m_2 = 1 M_\odot$ and $m_1$ ranging between $1 M\odot$ and $15 M_\odot$;
in the lower panel, we consider equal-mass binaries. We take two values of the opening angle; only the first object (``$m_1$'') is spinning (maximally). We notice that the number of precession cycles is larger for
binaries with asymmetric mass ratios than for equal-mass binaries, but
that the largest number of precession cycles occurs for binaries with small, comparable masses. For these, the BCV2 DTF has good performance: for instance, we find an average FF $\sim 0.965$ for $(2+2) M_\odot$ systems (in the search, we fix $f_{\rm cut} = 1000$~Hz).

\subsection{Performance of the BCV2P detection template family}
\label{sec:physDTF}

As we just saw, the BCV2 DTF offers good performance in matching the target
waveforms. Its simple form allows the formulation of a simple prescription for template bank placement (as we shall see in Sec.~\ref{sec:metric}), and the computational requirements are arguably economical. However, it is not straightforward to extract physical information from the BCV2 DTF parameters, which are phenomenological. In this section, we discuss a modification of the BCV2 DTF for single-spin target systems, which is also written in the frequency domain, where the phenomenological intrinsic parameters $\psi_0$, $\psi_{3/2}$, and $\mathcal{B}$ are replaced by the physical parameters $m_1$, $m_2$, $\chi$, and $\kappa$.

The first natural step toward a more physical parametrization is to replace the unmodulated phasing, $\psi_0 f^{-5/3} + \psi_{3/2} f^{-2/3}$, with the standard 2PN SPA phasing (as given, for instance, by Eq.\ (94) of Ref.\ \cite{bcv2}), which is a function of $m_1$, $m_2$, $\chi$, and $\kappa$. The second step is to replace the power-law precession angle $\alpha_p=\mathcal{B} f^{-2/3}$ with the analytic expression derived at Newtonian level Eq.\ \eqref{alphapk}, which is valid for single-spin systems, and which matches the numerical evolution of $\alpha_p$ quite well [see Fig.~(\ref{fig:alpha-p})]. In addition, since we expect this template family to be used in searches for asymmetric--mass-ratio binaries, for which the ending frequency falls at the margin of the band of good interferometer sensitivity, we do not include $f_\mathrm{cut}$ among the intrinsic parameter, but instead we fix it to the GW frequency of the ISCO, as evaluated in the test mass limit. We denote this modified frequency-domain DTF as BCV2P (where the P in BCV2P stands for both ``Pan'' and ``physical'').

We emphasize that, although physically parametrized, the BCV2P DTF is not on the average much closer to the target waveforms than the BCV2 DTF. Both DTFs suffer from the limitation emphasized in Sec.~\ref{subsub}: namely, that the frequency components that appear in the modulated GW phasing because of the evolution of the polarization tensors do not simply occur at the precession frequency $\Omega_p$, but also at its multiples (see Fig.~\ref{fig:fmodsig}). Thus, the structure of the precession-frequency harmonics and the amplitude modulations are not reproduced perfectly in either the BCV2 or the BCV2P DTFs. Truly exact physical templates for adiabatic spinning waveforms are so far available only in the time domain (where they are computed by solving the equations of motion) for single-spin binaries~\cite{pbcv1}.

\begin{figure}
\begin{center}
\includegraphics[width=0.42\textwidth]{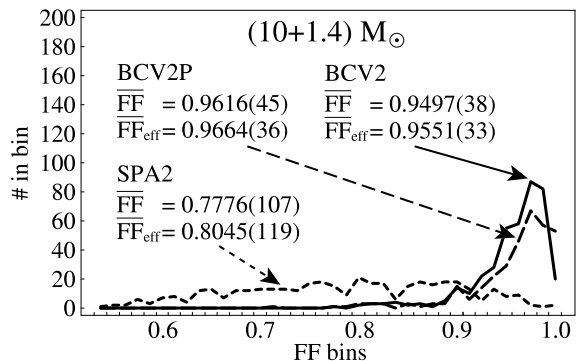}
\caption{Distribution of BCV2, BCV2P, and 2PN SPA fitting factors for $(10 + 1.4)M_\odot$ BH--NS target systems with uniformly sampled directional and local angular parameters (400 sets). See Fig.\ \ref{fig:FFs} for details, but note that here FF bins are shown on a linear scale.
\label{fig:FFs1014}}
\end{center}
\end{figure}

\begin{figure}
\begin{center}
\includegraphics[width=0.42\textwidth]{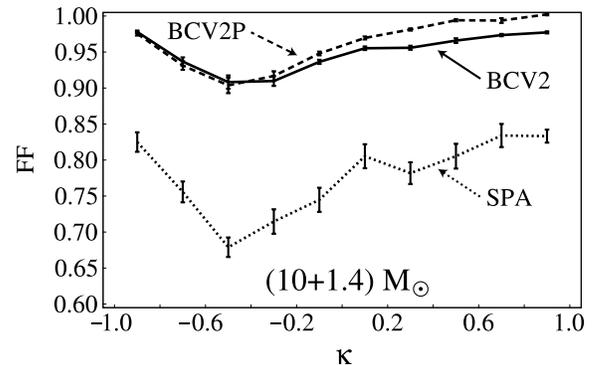}
\caption{Average fitting factors achieved by the BCV2, BCV2P, and 2PN SPA DTFs for $(10 + 1.4)M_\odot$ BH--NS target systems with uniformly sampled directional and local angular parameters (400 sets), plotted against $\kappa$. The vertices of the segmented curves show the FFs
averaged on the sets of target systems with $\kappa$ in the bins $[-1.0,-0.8),[-0.8,-0.6),\dots,[0.8,1.0]$. The bars show the sampling error on the bin averages.
\label{fig:FFvskappa}}
\end{center}
\end{figure}

We test the signal-matching performance of the BCV2P DTF only for $(10+1.4)M_{\odot}$ BH--NS binaries. The BCV2P DTF is as effective as the BCV2 DTF, and much better than the standard unmodulated stationary-phase--approximated templates (SPA). We plot the distributions of FFs for these three families in Fig.~\ref{fig:FFs1014}, which shows also the average and effective average FF. In Fig.~\ref{fig:FFvskappa}, we plot FF as a function of the target-system parameter $\kappa$.
 
In Tab.~\ref{tab:parest}, we examine the parameter-estimation capabilities of $M$, $\eta$, $\mathcal{M}$, $\chi$ and $\kappa$, by giving three characteristic quantities for each parameter:
\begin{itemize}
\item the bias, defined as the average systematic error of the FF projection (it might be possible to remove the bias partially by careful characterization of the projection map);
\item the rms systematic error, caused by the spread in the FF projection due to the presence of unmodeled target-system parameters;
\item the Gaussianity of the distribution, as characterized by the percentage of estimators falling in the 1-$\sigma$ and 3-$\sigma$ intervals (for a Gaussian distribution, these should be $\sim$ 69 \% and $\sim$ 100 \%, respectively).
\end{itemize}
These systematic errors are distinct from statistical error due to detection in noise, which is roughly inversely proportional to source strength (\emph{i.e.}, S/N). We see that chirp mass is the parameter that can be estimated most precisely, with $\sim 1\%$ bias and
rms deviation. Interestingly, $\kappa$ can also be estimated rather well.
\begin{table}
\begin{tabular}{c||r@{.}l|r@{.}l|c}
parameter & \multicolumn{2}{c|}{bias} & \multicolumn{2}{c|}{rms} & \% within 1-$\sigma$/3-$\sigma$ \\
\hline
$M$ & 10&4\% & 13&9\% & 82\%/98\% \\
$\eta$ & $-0$&104 (abs) & 0&187 (abs) & 78\%/98\% \\
$\mathcal{M}$ & 1&2\% & 1&3\% & 60\%/100\% \\
$\chi$ & $-20$&0\% & 15&1\% & 80\%/96\% \\
$\kappa$ & 0&020 (abs) & 0&153 (abs) & 80\%/98\%
\end{tabular}
\caption{
Bias, systematic rms error, and percentage of estimators falling in the 1-$\sigma$ and 3-$\sigma$ intervals for the BCV2P DTF parameters $M$, $\eta$, $\mathcal{M} = M \eta^{3/5}$ (the \emph{chirp mass}), $\chi$ and $\kappa$. The rms errors for $M$, $\mathcal{M}$, and $\chi$ are given as percentages of the target-system value of those parameters.
\label{tab:parest}}
\end{table}

Despite its fine parameter-estimation performance, the BCV2P DTF has some disadvantages with respect to the BCV2 DTF and to the time-domain physical templates of Ref.\ \cite{pbcv1}. For example, it will be more complex to build the template match metric and to place down templates for BCV2P than for BCV2 (see Sec.~\ref{sec:metric} below). From the point of view of FF, false-alarm rate, and parameter estimation, the BCV2P DTF is less attractive than the physical templates of Ref.\ \cite{pbcv2}. However, these might well be too computationally burdensome to be implemented for online searches; in that case, the BCV2P DTF could be use as an efficient first stage in a hierarchical search strategy.

\section{A procedure for template placement using the template match metric}

In this section we show how to place the BCV2 templates within a certain DTF parameter region, while guaranteeing a chosen \emph{minimum match} MM \cite{DIS,DA91} defined by
\begin{eqnarray}
\mbox{MM} &=& \min_{\lambda}\max_{\lambda' \in {\rm bank}} \rho(\lambda,\lambda') \nonumber \\
& \equiv & \min_{\lambda}\max_{\lambda' \in {\rm bank}} \frac{\langle h(\lambda) , h(\lambda') \rangle } {\sqrt{ \langle h(\lambda) , h(\lambda)  \rangle
 \langle h(\lambda') , h(\lambda')  \rangle } } \label{minmatch}
\end{eqnarray}
(where $\rho$ is the \emph{match}) in terms of the noise inner product 
\begin{equation}
\langle g, h \rangle \equiv  4 \, {\rm Re} \int_0^{+\infty} 
\frac{\tilde{g}^*(f) \tilde{h}(f)}{S_n(f)} df\,,
\end{equation}
with $S_n(f)$ the one-sided noise spectral density (given for this paper by Eq.\ (68) of Ref.~\cite{bcv1}). Although the maximization over the extrinsic DTF parameters can be carried out analytically, the existence of extrinsic parameters still influences the placement of templates, as discussed, \emph{e.g.}, in Sec.~VI of Ref.~\cite{pbcv1}. 

\subsection{Template metric of the BCV2 DTF}
\label{sec:metric}

The match between templates with close parameter values can be approximated using a \emph{metric} in parameter space~\cite{BSD,O,OS99},
\begin{equation}
\rho(\lambda,\lambda+\Delta\lambda)=1-g_{CD} \Delta\lambda^C \Delta\lambda^D\,.
\end{equation}
The components of the metric can be expressed in terms of first derivatives of the template waveforms,
\begin{multline}
\label{fisher}
g_{CD}=\frac{1}{2}
\left[
\frac{1}{\langle h , h \rangle}\Big\langle
\frac{\partial h}{\partial \lambda^C} ,
\frac{\partial h}{\partial \lambda^D},\rangle\right. \\
\left.-\frac{1}{\langle h | h \rangle^2}\Big\langle
\frac{\partial h}{\partial \lambda^C} ,
h\Big\rangle \Big\langle h ,\frac{\partial h}{\partial \lambda^D}
\Big\rangle \right]\,.
\end{multline}
It should be noted that the match between nearby templates with different cutoff frequencies cannot be described by the metric of Eq.\ \eqref{fisher}; in this paper, we shall consider only the problem of placing templates that share the same cutoff frequency $f_\mathrm{cut}$. For binaries with low masses, the waveforms end at relatively high frequencies compared to the band of good interferometer sensitivity, so $f_{\rm cut}$ does not play an important role. (For instance, adopting the language of Sec.\ VI of Ref.\ \cite{bcv1}, $f_{\rm cut}$ can be set by requesting that $\langle h(f_{\rm cut}),h(+\infty) \rangle \simeq 0.99$, which yields $f_{\rm cut} \simeq 400$ Hz using the Newtonian amplitude evolution $f^{-7/6}$.)

To evaluate the metric $g_{CD}$, we project $h(\psi_0,\psi_{3/2},{\mathcal B},f_{\rm cut},{\mathcal C}_k;f)$ [Eq.\ \eqref{bcv2}] onto an orthonormal basis $e_i(f)$ ($j=1,\ldots,6$), writing
\beq
\label{BCV2good}
h=A^i e_i(f), \qquad
\langle e_i , e_j\rangle =\delta_{ij}\,;
\eeq
the template is then normalized if
\beq
\label{hh}
\langle h,h \rangle = A_j A^j = 1\,.
\eeq
A convenient choice of the basis functions $e_i(f)$ is the following: obtain the functions $\widehat{e}_{1,3,5}(f)$ from the Schmidt orthonormalization procedure,
\beq
\left[\begin{array}{c} \widehat{e}_1 \\  \widehat{e}_3 \\ \widehat{e}_5 \end{array}\right]
=
\left[\begin{array}{ccc} a_{11} \\ a_{31} & a_{33}\\
a_{51} & a_{53} & a_{55}\end{array}\right]
\left[\begin{array}{c} 1 \\  \cos(\mathcal{B} f^{-2/3}) \\ \sin(\mathcal{B} f^{-2/3})
\end{array}\right] f^{-7/6}
\eeq
(note that the $a_{ij}$ are functions of $\mathcal{B}$); define $\widehat{e}_{2,4,6}(f)$ 
from
\beq
\widehat{e}_{n+1}(f) \equiv i \, \widehat{e}_{n}(f)\,,\qquad n=1,3,5\,;
\eeq
(for $f > 0$); finally, define
\beq
\label{enf}
e_n(f) = \widehat{e}_n(f) e^{i\Psi(f)+2 \pi i f t_c}\,.
\eeq
Our parameter set is now $\lambda^C\equiv\{A^i;x^\alpha \}\equiv\{A^i;t_c,\mathcal{B},\psi_0,\psi_{3/2}\}$,
with $i=1,\ldots,6$ and $\alpha=0,1,2,3$. The $A^i$, along with $x^0\equiv t_c$, are extrinsic parameters, while the $x^{\hat\alpha}$ (for $\hat\alpha=1,2,3$), are intrinsic parameters.

Equation \eqref{hh} is complemented by the useful relations
\begin{eqnarray}
\bigip{\pp{h}{A^i}}{h}&=&A_i \,, \label{relationsa} \\
\bigip{\pp{h}{x^\alpha}}{h}&=&0 \,, \\
\bigip{\pp{h}{A^i}}{\pp{h}{A^j}}&=&\delta_{ij} \,, \\
\bigip{\pp{h}{A^i}}{\pp{h}{x^\alpha}}&=&A^j \bigip{e_i}{\pp{e_j}{x^\alpha}} \,, \\
\bigip{\pp{h}{x^\alpha}}{\pp{h}{x^\beta}}&=& A_iA_j\bigip{\pp{e_i}{x^\alpha}}{\pp{e_j}{x^\beta}}\,; \label{relationsb} \\
\nonumber
\end{eqnarray}
also, the first derivatives of $e_n(f)$ with respect to the $x^\alpha$ can be summarized in the differential expression
\begin{multline}
\label{den}
\mathrm{d}e_n =
\left[
2\pi i f  \widehat{e}_n  \,\mathrm{d}t_c
+ i f^{-5/3} \widehat{e}_n  \,\mathrm{d}\psi_0\right. \\
\left. + i f^{-2/3} \widehat{e}_n  \,\mathrm{d}\psi_{3/2}
+ \frac{\partial\widehat{e}_n(f)}{\partial\mathcal{B}} \,\mathrm{d}\mathcal{B}
\right] e^{i\Psi(f)+2\pi i f t_c} \,.
\end{multline}
Assuming $A_i A^i=1$, we write the match between nearby templates as
\begin{widetext}
\beq
\label{fullmm}
\rho(\lambda,\lambda+\Delta\lambda)=1-g_{CD}(A^i,\mathcal{B})\Delta\lambda^C \Delta\lambda^D=1-
\frac{1}{2}
\left[
\begin{array}{cc}
\Delta A^i & \Delta x^\alpha
\end{array}
\right]
\left[
\begin{array}{c|c}
\displaystyle \delta_{ij}-A_i A_j &
\displaystyle A^l\bigip{e_i}{\pp{e_l}{x^\beta}}
\\
\\
\hline
\\
\displaystyle A^l\bigip{\pp{e_l}{x^\alpha}}{e_j} & \displaystyle A_l A_m \bigip{\pp{e_l}{x^\alpha}}{\pp{e_m}{x^\beta}}
\end{array}
\right]\left[
\begin{array}{c}
\Delta A^j \\ \\ \Delta x^\beta
\end{array}
\right]\,.
\eeq
\end{widetext}
In language of Ref.~\cite{pbcv1}, $g_{CD}$ is the \emph{full} metric, which describes the match between nearby templates in terms of differences between \emph{all} their parameters. In general, $g_{CD}$ can depend on all intrinsic and extrinsic parameters, but in our case it depends only on $A^i$ and $\mathcal{B}$, but not on $\psi_{0,3/2}$ and $t_c$ [as a consequence of Eqs.\ \eqref{enf}].

To determine the spacing of bank templates along the intrinsic-parameter directions, we work in terms of the extrinsic-parameter--maximized match
\begin{multline}
\rho_{\rm max}(A^j,x^{\gamma};x^{\hat\gamma}+\Delta x^{\hat\gamma}) \equiv \\
\equiv
\max_{\Delta A^j,\Delta t_c}\rho_{\rm max}(A^j,x^{\gamma};A^j+\Delta A^j,t_c+\Delta t_c,x^{\hat\gamma}+\Delta x^{\hat\gamma})\,,
\end{multline}
which is approximated by the \emph{projected} metric $g^{\rm proj}_{\hat\alpha \hat\beta}$, calculated by maximizing Eq.\ \eqref{fullmm} over $\Delta A^j$ and $\Delta t_c$, while fixing $\Delta x^{\hat\alpha}$ (see Eq.~(65) of Ref.~\cite{pbcv1}). Maximizing first over $\Delta A^j$, we notice that the submatrix $\delta_{ij}-A_i A_j$
is degenerate, with the single null eigenvector $A^i$ [this degeneracy occurs because the match \eqref{fullmm} remains constant when $\Delta A^i$ moves parallel to $A^i$---it must then be broken when we impose, without loss of generality, $A_k \Delta A^k=0$]. 
We find the maximum
\beq
\label{projAij}
1-\frac{1}{2}
\left[A^iA^j
G_{ij\alpha\beta}(\mathcal{B}) \right]\Delta x^\alpha \Delta x^\beta
\eeq
at the location
\beq
\Delta A^k = - A^j\bigip{e^k}{\pp{e_j}{x^\beta}}\Delta x^\beta
\eeq
[note that $A_k \Delta A^k=0$ due to Eq.\ \eqref{relationsa}]. The tensor $G_{ij\alpha\beta}$ of Eq.\ \eqref{projAij} is given by
\beq
G_{ij\alpha\beta}(\mathcal{B}) \equiv \bigip{\pp{e_i}{x^\alpha}}{\pp{e_j}{x^\beta}}
-\bigip{\pp{e_i}{x^\alpha}}{e_k}\bigip{e^k}{\pp{e_j}{x^\beta}}\,.
\eeq
Further maximizing \eqref{projAij} over $\Delta t_c$, we find the maximum
\begin{equation}
\rho_{\rm max}(A^j,x^{\gamma};x^{\hat\gamma}+\Delta x^{\hat\gamma}) = 
1 - g^{\rm proj}_{\hat{\alpha}\hat{\beta}}(A^j,\mathcal{B})\,\Delta x^{\hat\alpha}\Delta x^{\hat\beta}\,,
\end{equation}
where $g^{\rm proj}_{\hat{\alpha}\hat{\beta}}$ is the \emph{three-dimensional} projected metric
\begin{equation}
\label{projmetric}
g^{\rm proj}_{\hat{\alpha}\hat{\beta}}(A^j,\mathcal{B}) = \frac{1}{2}\left[A^i A^j G_{i j \hat\alpha \hat\beta}\right.
\left. -\frac{A^i A^j A^l A^m G_{ij 0\hat\alpha}
G_{lm 0\hat\beta}}{A^i A^jG_{ij00}}
\right]\,.\nonumber
\end{equation}
In general, the projected metric can depend on all the parameters, but in our case it depends only on $A^i$ and $\mathcal{B}$.

We can now go back to the problem of template placement. Within the metric approximation, if we choose a target template [the $h(\lambda)$ of Eq.\ \eqref{minmatch}] with parameters $\lambda^C \equiv (A^j,x^\gamma)$, the nearby templates with (extrinsic-parameter--maximized) match greater than MM must have intrinsic parameters $x^{\hat\gamma}+\Delta x^{\hat\gamma}$ that lie within the ellipse $\mathcal{E}(A^j,\mathcal{B},\mbox{MM})$ specified by 
\beq
\label{projcontour}
\mathcal{E}(A^j,\mathcal{B},\mbox{MM}): \; g^{\rm proj}_{\hat{\alpha}\hat{\beta}}(A^j,\mathcal{B})\,\Delta x^{\hat\alpha}\Delta x^{\hat\beta} \le  1-\mbox{MM}\,.
\eeq
The shape of this ellipse depends on the target-template extrinsic parameters $A^j$, which is not appropriate for the placement procedure that we are seeking, which should be formulated in terms of the intrinsic parameters only.

To that purpose, Ref.\ \cite{pbcv1} suggested adopting the \emph{average} shape of the match ellipse, as obtained by averaging the maximized match [and hence the left-hand side of Eq.~\eqref{projcontour}] over the extrinsic parameters. In Ref.\ \cite{pbcv1}, the averaging weights were determined from the prior distribution of the target-signal physical extrinsic parameters, in such a way that a template placement procedure guided by the average match contour would guarantee a certain \emph{expected} detection efficiency.
\begin{figure}[t]
\includegraphics[width=0.35\textwidth]{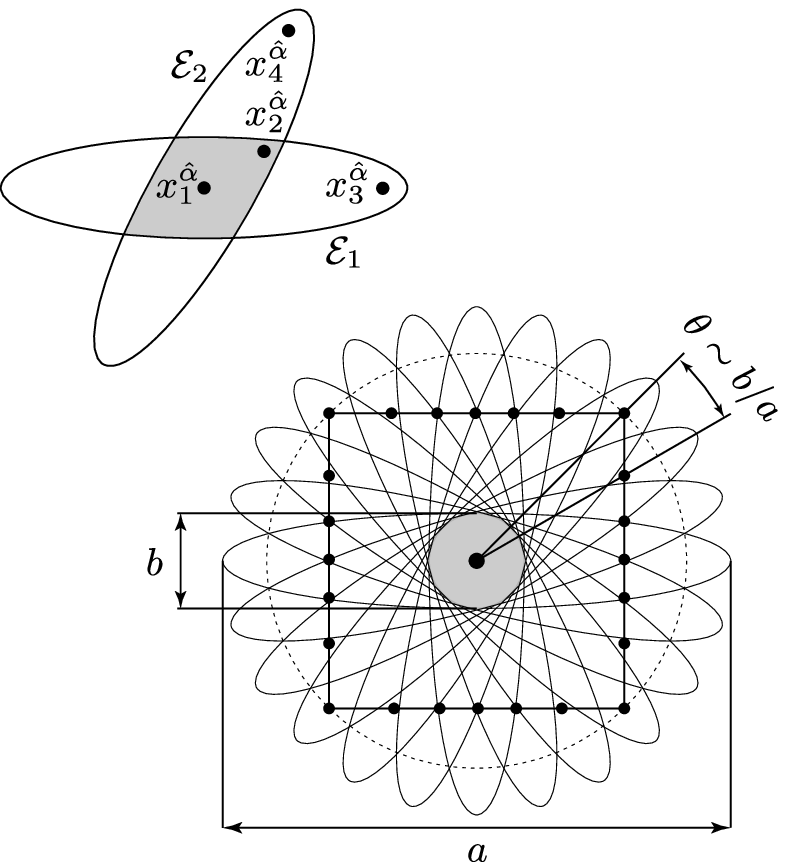}
\caption{Strategies to place templates in the space of BCV2 intrinsic parameters. \\
\textit{Top.} Suppose we have already placed a template at $x_1^{\hat\alpha}$, and consider only two sets of target-template extrinsic parameters, associated with which are two different match ellipses, $\mathcal{E}_1$ and $\mathcal{E}_2$. In order for the bank to guarantee the MM for both sets of extrinsic parameters, each of $\mathcal{E}_{1}$ and $\mathcal{E}_{2}$ must contain at least another template at a location different from $x_1^{\hat\alpha}$. This can be achieved with two templates ($x_{3}^{\hat\alpha}$ and $x_{4}^{\hat\alpha}$ in the figure), or more easily (but less optimally) with a single template ($x_2^{\hat\alpha}$ in the figure) in the intersection of $\mathcal{E}_{1}$ and $\mathcal{E}_{2}$. \\
\textit{Bottom.} In the idealized situation where all ellipses have the same elongated shape, but take all possible orientations, placing templates along a \emph{multilattice} (black dots) can be much more efficient than tiling on the basis of the minmax region (gray).
To obtain the multilattice, we construct a set of maximized-match ellipses evenly separated by angles $\theta \sim b/a$, so that the circle (dashed) through the intersections of the ellipses has radius $\sim a$. The unit cell of the multilattice is then given by the intersections of the semimajor axes of the ellipses with the square inscribed in the dashed circle.
\label{fig:extrinsic}}
\end{figure}

In our case, however, the extrinsic parameters are not physical, and do not have obvious prior distributions. We take a conservative approach, and we require that
for every value of the intrinsic parameters $x^{\hat\gamma}$ in the bank \emph{and} for every possible value of the extrinsic parameters $A^j$, there exist a nearby bank template with $\Delta x^{\hat\gamma}$ within the ellipse $\mathcal{E}(A^j,\mathcal{B},\mbox{MM})$. In principle, the neighboring template that satisfies the criterion could be different for different $A^j$, as illustrated at the top of Fig.\ \ref{fig:extrinsic}. For simplicity, however, we adopt a suboptimal strategy, requiring the existence of a nearby template within the \emph{intersection} of all the ellipses centered at $x^{\hat\gamma}$ (the gray region at the top of Fig.~\ref{fig:extrinsic}),
\beq
\mathcal{E}_{\rm minmax}(\mathcal{B},\mbox{MM}) \equiv \bigcap_{A^j}\mathcal{E}(A^j,\mathcal{B},\mbox{MM})\,.
\eeq
In analogy to the usage of Ref.\ \cite{DIS}, we denote $\mathcal{E}_{\rm minmax}(\mathcal{B},\mbox{MM})$ as the \emph{minmax} region, because it corresponds to considering the contours of the match maximized over the extrinsic parameters of search templates, and minimized over the extrinsic parameters of the prospective target signals,
\beq
\label{eq:minmax}
\min_{A^j}\left[ 1- g^{\rm proj}_{\hat{\alpha}\hat{\beta}}(A^j,\mathcal{B})\,\Delta x^{\hat\alpha}\Delta x^{\hat\beta}\right] \ge  \mbox{MM}\,.
\eeq
The minmax regions $\mathcal{E}_{\rm minmax}(\mathcal{B},\mbox{MM})$ may no longer be elliptical in shape, but their linear dimensions still scale linearly with $\sqrt{1-\mbox{MM}}$, as it happened for the ellipses $\mathcal{E}(A^j,\mathcal{B},\mbox{MM})$. 

It is worth pointing out that if the orientations of the ellipses $\mathcal{E}(A^j,\mathcal{B},\mbox{MM})$ vary substantially when $A^j$ is changed, relying on different nearby templates to achieve the MM for different extrinsic parameters can be much more efficient than relying on the minmax region. As a simple example, consider a situation in which all the ellipses $\mathcal{E}(A^j,\mathcal{B},\mbox{MM})$ have the same shape, with semi-major and semi-minor axes $a$ and $b$ (with $a \gg b$), but assume all possible orientations, as shown at the bottom of Fig.~\ref{fig:extrinsic}.  In this case, the minmax region has area $\sim b^2$. If we place templates according to the minmax prescription, the template density becomes $\sim 1/b^2$, much larger than the density $\sim 1/(ab)$ associated with each individual ellipse. 

On the other hand, we could place templates on the sides of squares with size $\sim a$, separated by a parameter distance $\sim b$, as shown at the bottom of Fig.~\ref{fig:extrinsic}. In this \emph{multilattice}, the average area occupied by each template is $ \sim a^2/(a/b) \sim ab$, corresponding to a density $\sim 1/(ab)$, much better than obtained with the minmax prescription.  Unfortunately, generating the appropriate multilattice is definitely more complicated than using the minmax prescription; it is also not clear whether for the BCV2 DTF we have in fact elongated ellipses with dramatically different orientations. Thus, we will adopt the minmax prescription in the rest of this paper. 
\begin{figure}[t]
\psfrag{ps0}[][]{$\Delta\psi_0$}
\psfrag{ps3}[][]{$\Delta\psi_{3/2}$}
\psfrag{b}[][]{$\Delta\mathcal{B}$}
\includegraphics[width=0.45\textwidth]{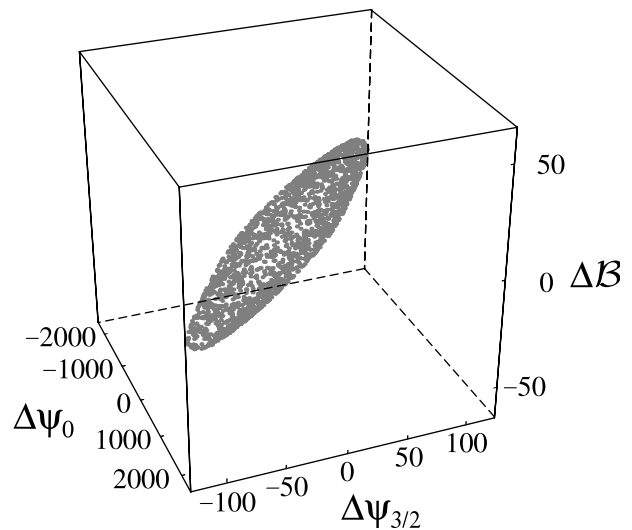}
\caption{The BCV2 DTF minmax region for $\mathcal{B}=60$, $f_{\rm cut}=400$ Hz (and any $\psi_0$, $\psi_{3/2}$), obtained by sampling Eq.\ \eqref{eq:minmax} with a random distribution of $A^i$.
\label{fig:contour}}
\end{figure}

We can approximate the minmax regions for the BCV2 DTF by sampling the $A^i$ randomly. In Fig.~\ref{fig:contour}, we plot an example minmax region for $\mathcal{B}=60$, $f_{\rm cut}=400$ Hz. This figure is typical in the sense that the minmax regions in the $(x^{\hat{1}},x^{\hat{2}},x^{\hat{3}})=(\psi_0,\psi_{3/2},\mathcal{B})$ space can be approximated rather well by ellipses. As a consequence, we can rely on (yet another) metric $\hat{g}_{\hat\alpha\hat\beta}(\mathcal{B})$ in the space of intrinsic parameters, whose match ellipses lie \emph{within} the corresponding minmax regions, but have similar volumes.  Once we are equipped with $\hat{g}_{\hat\alpha\hat\beta}(\mathcal{B})$, the placement of the BCV2 templates can be performed along the lines of conventional template placement procedures.

\subsection{Template placement}
\label{count}
\begin{figure}[t]
\includegraphics[width=0.45\textwidth]{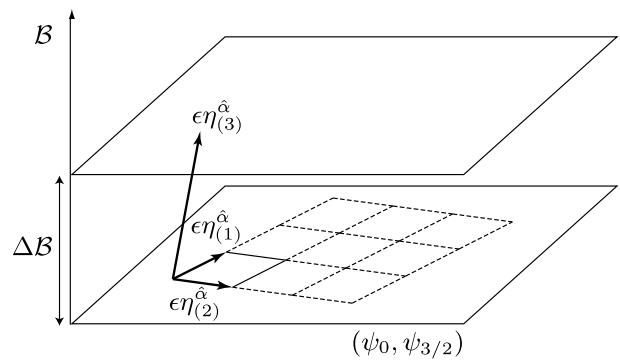}
\caption{Template placement in the BCV2 intrinsic-parameter space $(\psi_0,\psi_{3/2},\mathcal{B})$.
The orthonormal basis $(\eta^{\hat\alpha}_{(1)},\eta^{\hat\alpha}_{(2)},\eta^{\hat\alpha}_{(3)})$ is constructed in such a way that $\eta^{\hat\alpha}_{(1)}$ and $\eta^{\hat\alpha}_{(2)}$ both lie in the $(\psi_0,\psi_{3/2})$ plane, along which the projected metric is constant.  Unit cells constructed along these basis vectors, as shown in the figure, can be extended to the entire intrinsic-parameter space consistently. The coefficient $\epsilon$ is $\sqrt{4(1-{\rm MM})/3}$. 
\label{fig:placement}}
\end{figure}

The standard \emph{local} prescription for template placement is to follow
a cubic lattice\footnote{Under some circumstances, other lattices can provide better packing: for the BCV2 DTF, preliminary tests suggest that a tetrahedral lattice could reduce the number of templates by one fourth with respect to a cubic lattice.} constructed with three orthonormal basis vectors
$\{\eta_{(1)}^{\hat\alpha},\eta_{(2)}^{\hat\alpha},\eta_{(3)}^{\hat\alpha}\}$,
setting the side length [as measured with the metric $\hat{g}_{\hat\alpha\hat\beta}(\mathcal{B})$] of the unit cell equal to $\sqrt{4(1-\mathrm{MM})/3}$~\cite{O}.
For general metrics that depend on location in parameter space, such a local lattice cannot usually be extended consistently to cover the entire space. Luckily, this is possible in our case because of the translational invariance of $\hat{g}_{\hat\alpha\hat\beta}(\mathcal{B})$ along the $\psi_0$ and $\psi_{3/2}$ directions.

\begin{figure}[t]
\includegraphics[width=0.45\textwidth]{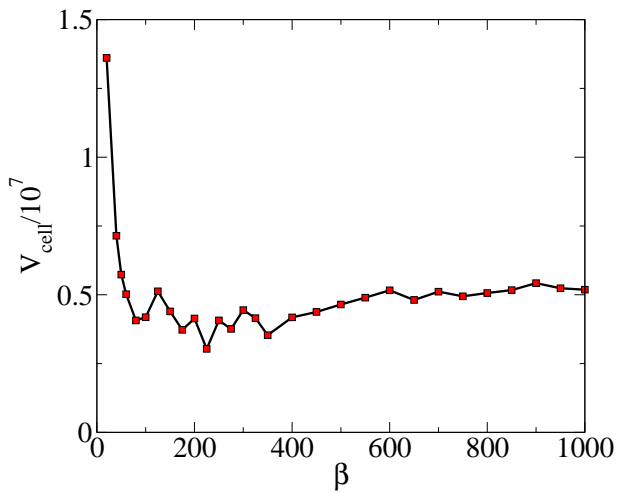}
\caption{Effective parameter volume of a single template cell as a function of $\mathcal{B}$, assuming a cubic lattice with MM = 0.97. We fix $f_{\rm cut}=400$ Hz.
\label{fig:size}}
\end{figure}

We first identify a set of orthogonal basis vectors $\{\eta_{(1)}^{\hat\alpha},\eta_{(2)}^{\hat\alpha},\eta_{(3)}^{\hat\alpha}\}$ at each point $x^{\hat \alpha}$, with the property that both $\eta_{(1)}^{\hat\alpha}$ and $\eta_{(2)}^{\hat\alpha}$ lie within the $(\psi_0,\psi_{3/2})$ plane. One such set follows from defining
\beq 
\eta_{(3)}^{\hat\alpha} \equiv
\frac{1}{\sqrt{\hat g^{\hat3 \hat3}}} g^{\hat\alpha\hat 3}\,; 
\eeq 
this $\eta_{(3)}^{\hat\alpha}$ is orthogonal to all tangent vectors that lie within the $(\psi_0,\psi_{3/2})$ plane. We can complete the basis with any pair of
$\hat{g}$-orthonormal vectors $\{\eta_{(1)}^{\hat\alpha},\eta_{(2)}^{\hat\alpha}\}$ in that plane.
Due to translational invariance, the cubic lattice constructed with this basis can be extended consistently along the $\psi_0$ and $\psi_{3/2}$ directions, covering a thin slice of parameter space parallel to the $(\psi_0,\psi_{3/2})$ plane, with coordinate thickness
\beq
\Delta \mathcal{B} = \sqrt{\frac{4(1-\mathrm{MM})}{3}}
\eta_{(3)}^{\hat 3} = \sqrt{\frac{4(1-\mathrm{MM})}{3}} \sqrt{\hat
g^{\hat3 \hat3}}
\eeq
along the $\mathcal{B}$ direction. We can stack these slices to cover the entire tridimensional parameter space (see Fig.~\ref{fig:placement}). In Fig.~\ref{fig:size}, we plot the parameter volume of the cube inscribed in the MM=0.97 minmax regions (\emph{i.e.}, the effective volume around each template in the cubic lattice) as a function of $\mathcal{B}$.
 
We are now ready to give a rough estimate of the number of templates required for a matched-filtering search for BH--NS and BH--BH target systems. We assume a single value (400 Hz) for $f_{\rm cut}$. For BH--BH and BH--NS systems with component masses $(m_1,m_2) \in [6,12]M_\odot \times [1,3]M_\odot$, we select (following Fig.\ \ref{fig:projections}) the BCV2 DTF parameter region
\begin{equation}
\left[\frac{\psi_0}{10^5},\frac{\psi_{3/2}}{10^3},\frac{\mathcal{B}}{10^2} \right]
\in [1.5 , 8.5] \times
[-4.5 ,0.5]\times[0,10]\,;
\end{equation}
our estimate for the number of templates is $\mathcal{N}_{\mbox{\tiny BH--NS}} \simeq 7\times 10^5$ for MM=0.97. By contrast, for BH--BH systems with component masses $(m_1,m_2) \in [5,20]M_\odot \times [5,20]M_\odot$, we select the BCV2 DTF parameter region
\begin{equation}
\left[\frac{\psi_0}{10^5},\frac{\psi_{3/2}}{10^3},\frac{\mathcal{B}}{10^2} \right]
\in [0, 2.5] \times
[-2.5, 1]\times[0,4.5]\,;
\end{equation}
out estimate for the number of templates is $\mathcal{N}_{\mbox{\tiny BH--BH}} \simeq 8\times 10^4$ for MM=0.97. However, it is not clear that precessing-binary templates are needed for this entire region: for binaries with relatively high component masses, ``BCV1'' (unmodulated) templates can already yield high FFs.

\section{Conclusions}
\label{conc}

Reference \cite{bcv2} introduced the BCV2 DTF for use in precessing-binary searches; the BCV2 templates are written directly in the frequency domain, and depend on ten phenomenological parameters (the intrinsic parameters $\psi_0$, $\psi_{3/2}$, $\mathcal{B}$, and $f_{\rm cut}$, and the extrinsic parameters $\mathcal{C}_{1,\ldots,6}$). The modulational effects of precession in the target systems are modeled by the $\mathcal{C}_{1,\ldots,6}$ and by the single intrinsic parameter $\mathcal{B}$; this marks a definite improvement over previous attempts at spinning-binary templates \cite{ACST94,apostolatos0,apostolatos1,apostolatos2,GKV,GK,Gpc}, which require several intrinsic parameters to model precession, leading to larger and more computationally expensive search template banks.

In this paper, we have tested the signal-matching performance of the BCV2 DTF against target signals from asymmetric--mass-ratio binaries with component masses $(m_1,m_2) \in [6,12]M_\odot \times [1,3]M_\odot$ (with both components maximally spinning), and from $(10 + 1.4) M_\odot$ BH--NS binaries (with maximally spinning BHs and nonspinning NSs). The waveforms were computed at the 2PN order in the adiabatic approximation~\cite{2PN}. We found very good fitting factors, averaging between 0.94 and 0.98; our results are summarized in Fig.~\ref{fig:FFs}. By means of the FF projection map (see Sec.\ \ref{sec:nonphysDTF}) we also identified the region in BCV2 DTF parameter space that must be included in a template-bank--based search for these systems; our results are shown in Figs.\ \ref{fig:projections} and \ref{fig:psi0fcut}.

Because all the asymmetric--mass-ratio binaries considered in this paper generate waveforms with ending frequencies (\emph{i.e.}, MECO frequencies, see Eqs.\ (11)--(12) of Ref.\ \cite{bcv2}) at the margin of the band of good interferometer sensitivity, we are free to fix $f_\mathrm{cut}$ to a reasonable value (see, for instance, the discussion below Eq.\ \eqref{fisher}), reducing the dimensionality of template space without a corresponding degradation in the FFs.

In addition, by a closer study of precessional dynamics and GW generation in single-spin binaries, we were able to relate the BCV2 phenomenological parameter $\mathcal{B}$ to the physical parameters of the target binary (see Sec.\ \ref{sec:beta}). In the process of doing so, we realized that waveform modulations occur at the fundamental \emph{and} at higher and lower harmonics of the precession frequency $\Omega_p$ (for our target systems, the frequency at which the orbital angular momentum and the spin precess around the total angular momentum; for the BCV2 templates, the equivalent precession frequency corresponding to a choice of $\mathcal{B}$). In the target signals, the higher harmonics arise because modulation is caused by oscillations in the components of the polarization tensor [see Eq.\ \eqref{ee}], and not directly by the precession of the orbital angular momentum and spins. In the BCV2 templates, the higher harmonics follow naturally from Eq.\ \eqref{bcv2}, and can even be made dominant by an appropriate choice of the phenomenological coefficients
${\cal C}_{1,\ldots,6}$ (see Fig.\ \ref{fig:fmod}).

These considerations allowed us to understand certain features in the distribution of best-fit $\mathcal{B}$ against the target-system opening angle $\kappa$ that had remained unexplained in Ref.\ \cite{bcv2}. The analysis performed in this paper suggests also a modification of the BCV2 DTF, whereby the three phenomenological parameters $\psi_0$, $\psi_{3/2}$, and $\mathcal{B}$ are replaced by the four physical parameters $M$, $\eta$, $\kappa$ and $\chi$. This modified DTF (BCV2P) has a signal-matching performance comparable to (or slightly better than) the BCV2 DTF, for $(10 + 1.4) M_\odot$ BH--NS binaries (see Figs.~\ref{fig:FFs1014} and \ref{fig:FFvskappa}); we expect that we would find slightly higher FFs if we were to include higher-order PN corrections in $\alpha_p$. The BCV2P DTF has the advantage of providing the straightforward and reliable estimation of certain parameters of the target system, such as $\mathcal{M}$ and $\kappa$ (see Table~\ref{tab:parest}); it has the drawback of depending on four rather than three intrinsic parameters, and its template match metric would be even more complicated than the BCV2 metric described in Sec.\ \ref{sec:metric}.
Therefore, we suggest that the BCV2P DTF could be used for follow-up searches on a reduced set of BCV2 triggers, or to estimate target-system parameters, if it turns out that we cannot computationally afford the physical template family of Ref.\ \cite{pbcv1}, which is also written in terms of the physical parameters of the target system, and which yields FFs very close to unity.

Last, we computed the full ten-dimensional template match metric in the ($\psi_0$, $\psi_{3/2}$, $t_c$, $\mathcal{B}$, ${\cal C}_1, \ldots, {\cal C}_6$) parameter space,
and the tridimensional projected metric in the ($\psi_0$, $\psi_{3/2}$, $\mathcal{B}$) subspace, obtained by projecting out the seven extrinsic parameters. The projected metric does not depend on $\psi_0$ and $\psi_{3/2}$, but only on $\mathcal{B}$ (and $f_{\rm cut}$). We described a prescription to place BCV2 templates in a search bank using \emph{minmax regions}, and exploiting the ($\psi_0$, $\psi_{3/2}$) translation invariance of the metric. Fixing $f_{\rm cut} = 400$ Hz and adopting a cubic lattice, we find that the estimated number of templates required for MM = 0.97 is $\sim 7\times 10^5$ for target systems with component masses $(m_1,m_2) \in [6,12]M_\odot \times [1,3]M_\odot$, and $\sim 8\times 10^4$ for target systems with component masses $(m_1,m_2) \in [5,20]M_\odot \times [5,20]M_\odot$.
These numbers make it unlikely that a straightforward BCV2 search could be performed \emph{online}, with the current computational resources, for these full parameter ranges. Workarounds might include carefully isolating (by extensive Monte Carlo runs) the subregions of parameter space where spin effects are weaker, so that nonspinning templates (such as the BCV1 DTF \cite{bcv1}) can be substituted effectively for BCV2 templates, or employing the BCV1 and BCV2 DTFs sequentially in a hierarchical search.

\acknowledgments
We thank B.S.\ Sathyaprakash for useful discussions and the internal LSC referee for his valuable comments. Research led to this paper has been supported by NSF grant PHY-0099568. Y.C.'s research was also supported by the David and Barbara Groce Fund at the San Diego Foundation, as well as the Alexander von Humboldt Foundation's Sofja Kovalevskaja Award (funded by the German Federal Ministry of Education and Research); M.V.'s research was also supported by the LISA Mission Science Office at the Jet Propulsion Laboratory, California Institute of Technology, where it was performed under contract with the National Aeronautics and Space Administration.


\end{document}